\documentclass[11pt,a4paper]{article} 
\usepackage{jheppub}
\usepackage{overpic,sansmath}

\makeatletter\gdef\@fpheader{~}\gdef\@journal{}\makeatother

\begin{document}

\title{Derivative couplings in gravitational production in the early universe}

\author[a]{Daniel E. Borrajo Gutiérrez,}
\author[a]{Jose A. R. Cembranos,} 
\author[a,b]{Luis J. Garay,}
\author[a]{and Jose M. S\'anchez Vel\'azquez}

\affiliation[a]{Departamento de F\'isica Te\'orica \& IPARCOS, Universidad Complutense de Madrid, \\ 28040 Madrid, Spain}
\affiliation[b]{Instituto de Estructura de la Materia (IEM-CSIC), C/ Serrano 121, 28006 Madrid, Spain}

\emailAdd{dborrajo@ucm.es}
\emailAdd{cembra@ucm.es}
\emailAdd{luisj.garay@ucm.es}
\emailAdd{jmsvelazquez@ucm.es}

\abstract{
Gravitational particle production  in the early universe is due to the coupling of matter fields to curvature. This coupling may include derivative   terms that modify the kinetic term.  The most general first order  action  contains derivative couplings to the curvature scalar and to the traceless Ricci tensor, which can be dominant in the case of (pseudo-)Nambu-Goldstone bosons or disformal scalars, such as branons. 
In the presence of these derivative couplings, the density of produced particles for the adiabatic regime in the de Sitter phase (which mimics inflation) is constant in time  and  decays with the inverse effective mass (which in turn depends on the coupling to the curvature scalar). In the reheating phase following inflation,  the presence of derivative couplings to the background curvature modifies in a  nontrivial way the gravitational production even in the perturbative regime. We also show that the two couplings ---to the curvature scalar and to the traceless Ricci tensor--- are drastically different, specially for large masses. In this regime, the production becomes highly sensitive to the former coupling while it becomes independent of the latter.}

\maketitle

\section{Introduction}

Our current understanding of cosmology is based on the $\Lambda$CDM model, which introduces a dark matter sector to explain the dynamics of the Universe. This dark matter sector is yet poorly understood as the fundamental nature of its constituents remains still a mystery. Due to the seemingly negligible interactions between dark matter and the Standard Model (SM) sector \cite{Kahlhoefer2017,Gaskins:2016cha}, many authors have proposed in the last decades that dark matter may have been produced gravitationally during the early stages of the Universe \cite{Chung:1998zb, Chung:2001cb, Ema2018, Markkanen:2015xuw,Velazquez:2019mpj, Herring:2019hbe}. Gravitational production has been studied since the pioneer work of Parker \cite{Parker:1969au} in which the effect of spontaneous particle production by the expansion of the Universe was first studied. In general, we know now that particle creation is a general feature of quantum field theory in curved spacetimes which is the paradigm we use to study the early epochs of the Universe.

Inflation \cite{PhysRevD.23.347,LINDE1983177} provides the seeds for structure formation giving a nearly scale-invariant power spectrum of fluctuations due to the quantum fluctuations of the inflaton field \cite{Liddle, Mukhanov}. The bridge between inflation and the usual hot big bang scenario is obtained by an intermediate phase known as reheating in which the energy stored in the inflaton field is transmitted to the Standard Model degrees of freedom. Particle production due to the expansion of the Universe is specially relevant during these two stages of the Universe and the transition between them. The density of created particles during these primordial cosmological epochs could potentially explain the dark matter abundance that is currently observed.

With this goal in mind, we will consider a relativistic massive scalar field nontrivially coupled to gravity. In particular, we will include terms involving the field $\varphi$, its derivatives $\varphi_{,\mu}$, the Ricci tensor $R_{\mu\nu}$, and the scalar curvature $R$ in the action. The $R\varphi^2$ term has been largely studied in the previous literature \cite{Chung:1998zb, Chung:2001cb, Ema2018, Markkanen:2015xuw, Velazquez:2019mpj, Herring:2019hbe}, but terms containing derivatives of the field such as $\varphi^{,\mu}\varphi^{,\nu}R_{\mu\nu}$ and $\varphi^{,\mu}\varphi_{,\mu}R$ have not been studied in this gravitational production context as far as we know. 
The main goal of this work is to study particle production in the early universe in the presence of these derivative couplings. The existence of nonsuppressed nonminimal couplings of this kind appear in the context of (pseudo-)Nambu-Goldstone bosons associated with the breaking of global symmetries \cite{Nambu:1960tm,Goldstone:1961eq}.They are also generic for disformal scalar fields~\cite{Bekenstein:1992pj}, that are coupled to the Einstein tensor. An example of this type of fields is provided by branons in flexible brane-world models \cite{Dobado:2000gr,Cembranos:2001rp,Alcaraz:2002iu,Cembranos:2004eb,Cembranos:2016jun}. The phenomenology of such couplings have been extensively studied in colliders \cite{Alcaraz:2002iu,Brax:2014vva,Achard:2004uu,Cembranos:2011cm,Cembranos:2005sr,Cembranos:2005jc,Cembranos:2011cm}, with precision observables \cite{Cembranos:2005sr,Cembranos:2005jc}, and with astrophysical observations \cite{Cembranos:2003mr,Cembranos:2003fu,Maroto:2003gm,Cembranos:2011cm}.

This manuscript is organised as follows. In section \ref{creationsec} we introduce the model of a massive scalar field with a linear coupling to the Ricci scalar and linear derivative couplings to both the Ricci tensor and scalar, neglecting the back-reaction of the field on the background and the interaction with the SM particles. We then obtain the equation of motion for this field in a Friedmann-Lema\^{i}tre-Robertson-Walker background. We  quantise it and  briefly describe  the formalism of particle production in such curved spacetimes. In section \ref{dSsec} we solve this equation for de Sitter geometry as a special case to see how the derivative couplings affect particle production. As we will see, in this case the addition of derivative couplings does not alter the type  of equation for the modes of the field, simply introducing an effective mass which cannot be distinguished experimentally from the physical mass. In section \ref{multistagesec} we investigate the gravitational effects in a simplified model for the reheating background, allowing us to delve into the effects of the derivative couplings. The investigation in the reheating epoch is carried out numerically and we use an interpolation regime  for the transition between the de Sitter phase and the reheating one. We conclude in section \ref{conclusions} with a summary of the main results obtained in this work.

Notation: We use units such that $c=\hbar = 1$ and signature $(-+++)$.

\section{Particle creation in curved spacetimes}\label{creationsec}

Let us consider a massive scalar field with several direct linear couplings to the background curvature. The dynamics of this field in a given spacetime is encoded in the action
\begin{equation}\label{eq: action}
    \mathcal{S} = -\frac{1}{2}\int d^4x\sqrt{-g} \big[\varphi_{,\mu}\varphi^{,\mu}+(m^2+\xi R)\varphi^2\big]+\mathcal S_\text{der},
\end{equation}
where $g$ is the determinant of the spacetime metric $g_{\mu\nu}$, $R$ is the Ricci curvature scalar,  $\xi$ is a dimensionless constant, $m$ is the mass of the field $\varphi$, and $\mathcal S_\text{der}$ will be defined shortly. Apart from this latter term this is the well-known action for a nonminimally coupled scalar field without modifying the kinetic term. The derivative-coupling contribution to the action
\begin{equation}
   \mathcal{S}_\text{disf} = \frac{1}{2}\int d^4x\sqrt{-g}    \Big[\gamma R g^{\mu\nu}+\sigma\big( R^{\mu\nu} -\frac14 R g^{\mu\nu}\big) \Big]\varphi_{,\mu}\varphi_{,\nu},
\end{equation}
where
$R^{\mu\nu}$ is the Ricci tensor 
and $\gamma$ and $\sigma$ are parameters with units of inverse energy squared, has been written in this particular form in order to separate the part that describes the coupling to the curvature scalar from the part that contains the coupling to the traceless Ricci tensor. In fact, as we will see, these two contributions play an entirely different role in the particle production.

In cosmological scenarios the background geometry is well described, in average, by the Friedmann-Lema\^{i}tre-Robertson-Walker metric. For simplicity and in agreement with the current cosmological observations \cite{Ade:2015xua}, we are considering flat spatial sections. Moreover, for later convenience, we will write the FLRW metric using conformal time:
\begin{equation}
    ds^2 = a(\eta)^2(-d\eta^2 +\delta_{ij}x^{i}x^{j}).
    \label{eq:flrw}
\end{equation}
For this specific geometry, the Ricci tensor is diagonal and its spatial components, due to homogeneity and isotropy, are all equal. It can be expressed in terms of the scale factor $a(\eta)$,  the Hubble parameter $H=a'/a^2$, and the curvature scalar $R  = 6 a''/a^3$ as 
\begin{equation}\label{eq: curvature}
        R_{00}  = (3 H^2-R/2)a^2 , \qquad
        R_{ii}  = (H^2+R/6)a^2 ,
     \end{equation}
where the prime denotes derivative with respect to the conformal time $\eta$.

From the action \eqref{eq: action} we obtain the equation of motion for the field $\varphi$:
\begin{equation}
    (1+A)\varphi'' - (1+B)\nabla^2\varphi + C\varphi' + D\varphi = 0,
\end{equation}
where, for convenience, we have defined the functions
\begin{align}\label{functions} 
     A &=  3\sigma H^2 - (\gamma+\sigma/4) R,  &
         C &= 2aH(1 +E), \qquad E= A  + A'/(2aH), \nonumber\\
        B &= -\sigma H^2 + (\sigma/12-\gamma)R, &
        D &= (m^2+\xi R)a^2.
\end{align}
An appropriate rescaling of the field variable allows us to eliminate the friction term. Indeed, in terms of the rescaled field $\chi(\mathbf{x},\eta)=f(\eta)\varphi(\mathbf{x},\eta)$ with 
\begin{equation}
    f(\eta) = a(\eta)\sqrt{1 + A(\eta)},
\end{equation}
the equation of motion for  $\chi$ reads
\begin{equation}\label{chiNMDC}
    \chi'' - \frac{1+B}{1+A}\nabla^2\chi +  F\chi = 0,
\end{equation}
where
\begin{equation}\label{coefF}
   F = \frac{(1+A)(4D-2C')+2CA'-C^2}{4(1+A)^2}.  
\end{equation}
 
As a general comment, from \eqref{functions} and \eqref{coefF} we see that some terms of the field equation vanish if $\sigma=-4\gamma$, including a term proportional to the second derivative  of the curvature scalar $R''$. In this case the field couples only to the Einstein tensor, which is also the case for   disformal scalar fields and branons \cite{Bekenstein:1992pj,Dobado:2000gr,Cembranos:2001rp,Alcaraz:2002iu,Cembranos:2004eb,Cembranos:2016jun}.
Moreover, additional significant simplifications may appear for specific functional forms of the scale factor. For instance, in a  radiation-dominated phase ($a \propto \eta$), the curvature scalar vanishes; and   in a de Sitter phase, which mimics the dynamics of inflation (\mbox{$a \propto \eta^{-1}$}), all the curvature terms are proportional to the (constant) Hubble parameter in this phase.

Exploiting the fact that the spatial sections are flat, we can expand the field $\chi$ in Fourier amplitudes 
\begin{equation}
    \chi_{{\mathbf{k}}}(\eta) = \frac{1}{(2\pi)^{3/2}}\int\mathrm{d}^3\mathbf{x}\,\chi(\mathbf{x},\eta)e^{-i{\mathbf{k}}\cdot\mathbf{x}},
\end{equation}
which satisfy harmonic oscillator-like equations of motion
\begin{equation}\label{eqmodes}
    \chi_{\mathbf{k}}'' + \omega_k^2(\eta)\chi_{\mathbf{k}} = 0,
\end{equation}
where $k=|{\mathbf{k}}|$, with time-dependent frequency
\begin{equation}
    \omega_k(\eta) = \sqrt{\frac{1+B(\eta)}{1+A(\eta)}k^2 +  F(\eta)}.
    \label{eq:freqbare}
\end{equation}

We can write any solution as a linear combination of two complex conjugate solutions
$\{v_k(\eta), v_k^*(\eta)\}$. Due to the isotropy of the background geometry, we can choose bases of solutions which only depend  on $k$. Then the Fourier amplitude $\chi_{\mathbf{k}}$ can be expressed as a linear combination of the mode function $v_k$ and its complex conjugate, the coefficients being the creation and annihilation variables $a_{\mathbf{k}}$ and $a^*_{-{\mathbf{k}}}$:
\begin{equation}\label{eq: Fourier amplitude}
	\chi_{\mathbf{k}}(\eta) = \frac1{\sqrt{2}}\big[a_{\mathbf{k}} v_k(\eta) + a_{-{\mathbf{k}}}^* v_k(\eta)^* \big].
\end{equation}
In order to preserve the standard Poisson bracket structure for the annihilation and creation variables, the mode functions must be normalised. We will follow  the convention of Ref.~\cite{Mukhanov}:
\begin{equation}
    	W[v_k, v_k^*] = v_k' v_k^* - v_k (v_k^*)'= 2 i.
    	\label{eq:wronskian}
\end{equation}

The Fock quantisation procedure (see e.g. Ref. \cite{Mukhanov:2005sc}) can be carried out by promoting the annihilation and creation variables $a_{\mathbf{k}}$ and $a_{\mathbf{k}}^*$ to creation and annihilation operators $\hat a_{\mathbf{k}}$ and $\hat a_{\mathbf{k}}^\dagger$ satisfying the canonical commutation relations
\begin{equation}
    [\hat a_{\mathbf{k}}, \hat a_{{\mathbf{k}}'}^\dagger] =\delta({\mathbf{k}}-{\mathbf{k}}'), \qquad [\hat a_{\mathbf{k}}, a_{{\mathbf{k}}'}] =[\hat a_{\mathbf{k}}^\dagger, \hat a_{{\mathbf{k}}'}^\dagger]=0.
\end{equation}
The Hilbert space for the states of the field is spanned in the usual way by the action of finite linear combinations of products of these operators on the vacuum state, defined as the one which is annihilated by all the annihilation operators, i.e.,
\begin{equation}\label{vacuum}
	\hat a_{\mathbf{k}}\left|{0}\right> = 0, \qquad \forall\, {\mathbf{k}}.
\end{equation}

Generally speaking, in curved spacetimes, the definition of the vacuum state for the field is not unique but suffers from different ambiguities that can lead to potentially inequivalent quantum theories. Different choices of mode functions in \eqref{eq: Fourier amplitude} will define a different set of creation and annihilation operators and hence, different vacuum states. As each basis of mode functions is complete, different bases must be related by a linear transformation, i.e.,  if we consider a different set of mode functions $\{u_k, u^*_k\}$, we can write
\begin{equation}\label{cambiobase}
	v_k(\eta) = \alpha_k u_k(\eta) + \beta_k u_k^*(\eta),
\end{equation}
with $\alpha_k,\ \beta_k$ complex coefficients satisfying appropriate normalisation conditions that ensure the normalisation of the new modes. This leads to new creation and annihilation operators
related to   the  previous ones by the so-called Bogoliubov transformations, \begin{equation}\label{Bogoliubov}
 		\hat{b}^\dagger_{\mathbf{k}}  = \alpha_k\hat{a}_{\mathbf{k}}^\dagger + \beta_k^*\hat{a}_{-{\mathbf{k}}}, \qquad
		\hat{b}_{\mathbf{k}}  = \alpha_k^*\hat{a}_{\mathbf{k}} + \beta_k\hat{a}_{-{\mathbf{k}}}^\dagger.
\end{equation}
The vacuum states corresponding to each quantisation $\left|0\right>_a$ and $\left|0\right>_b$ are related by
\begin{equation}\label{bog}
	 |0\rangle_b = \left[\prod_{\mathbf{k}}{\frac1{|\alpha_k|^{1/2}}}\exp\left(-\frac{\beta_k^*}{2\alpha_k}\hat{a}_{\mathbf{k}}^\dagger
	\hat{a}_{-{\mathbf{k}}}^\dagger\right)\right] |0\rangle_a,
\end{equation}
provided that both quantisations are unitarily equivalent \cite{Shale, HoneggerRieckers, Ruijsenaars}, that is to say, if and only if  \mbox{$\sum_{\mathbf{k}}|\beta_k|^2<\infty$}. 

The Bogoliubov coefficients $\alpha_k$ and $\beta_k$ can be straightforwardly computed from \eqref{cambiobase} imposing the Wronskian condition \eqref{eq:wronskian}:
\begin{equation}\label{coefs}
    \alpha_k = \frac{ u^{\prime *}_k v_k -u^*_k v'_k}{2i}  , \qquad \beta_k =  \frac{u_k v'_k -u'_k v_k}{2i}.
\end{equation}
If we introduce the particle number operator $\hat N^{(a)}=\sum_{\mathbf{k}} \hat{a}_{\mathbf{k}}^\dagger\hat{a}_{\mathbf{k}}$ (and likewise for the  quantisation $(b)$), it is easy to see that the total number of particles $ {}_a\langle 0|\hat N^{(b)}|0\rangle_a$ corresponding to the quantisation $(b)$ in the vacuum $|0\rangle_a$, is divergent because the spatial volume is infinite. Introducing the particle-number density $\hat n^{(a)}$ to avoid this problem, it can be shown that
\begin{equation}
    {}_a\langle 0|\hat n^{(b)}|0\rangle_a=\sum_{\mathbf{k}} |\beta_k|^2.
    \label{betados}
\end{equation}

In cosmology, the gravitational production occurs because of the time evolution of the background geometry. In these scenarios, the vacuum state for a quantum field is not stationary. Particles are produced because the evolved vacuum state  at a given instant of time is in general  an excited state with respect to the instantaneous vacuum defined at that instant of time.

We can define the vacuum state of the scalar field through the initial conditions given in the asymptotic past. These initial conditions define a set of modes $v_k$ and $v^*_k$ to be discussed in the following sections. 
On the other hand, a comoving observer will have a different instantaneous notion of vacuum. We can use the so-called adiabatic prescription~\cite{BirrelDavies} to define this instantaneous vacuum. This prescription is appropriate and  well defined if the geometry evolves slowly enough as compared with the characteristic time scales of the field, i.e., whenever the frequency  $\omega_k(\eta)$ 
is a slowly changing function during the time $\Delta\eta = 1/\omega_k$ (adiabatic condition):
\begin{equation}\label{adiabatic}
	\left| \frac{\omega_k(\eta+\Delta\eta)-\omega_k(\eta)}{\omega_k(\eta)} \right| \approx \left|\frac{\omega_k'}{\omega_k}\Delta\eta\right|
	=   \left|\frac{\omega'_k}{\omega_k^2}\right| \ll 1.
\end{equation}
The corresponding modes are
then given by the WBK approximation to first order \cite{BirrelDavies} at a given $\eta_0$, which satisfy the following initial conditions:
\begin{align}\label{adiabaticos}
    u_k(\eta_0)  = \frac{1}{\sqrt{\omega_k(\eta_0)}},
    \qquad u_k'(\eta_0)  = \frac{-1}{\sqrt{\omega_k(\eta_0)}}\left(i\omega_k(\eta_0)+\frac{\omega_k'(\eta_0)}
    {2\omega_k(\eta_0)}\right).
\end{align} 

\section{Particle creation in de Sitter spacetime}\label{dSsec}

In order to explain the current cosmological observations with the standard cosmological model, people usually introduce an inflationary epoch prior to the usual big bang model.  For our purposes, we can consider this inflationary epoch as a pure de Sitter geometry, which can be described with a scale factor $a(\eta) = -1/({H_0\eta})$. Here the Hubble parameter $H_0$ is constant and the conformal time $\eta$ lies on the interval $(-\infty,0)$. In more realistic scenarios, the background geometry is governed by a slowly varying Hubble parameter in a quasi-de Sitter expansion.

As the de Sitter geometry is a maximally symmetric solution, the Ricci scalar is proportional to the metric, $R_{\mu\nu} \propto R g_{\mu\nu}$. Therefore, in this background, the coupling to the derivatives of the field depends only on the single parameter  $\gamma$. This is not surprising because for the de Sitter geometry the traceless Ricci tensor is identically zero.  Computing the curvature related quantities for this scale factor and substituting them in \eqref{eq: curvature} and \eqref{eq:freqbare}, we obtain the time-dependent frequency for the field during the inflationary phase:
\begin{equation}\label{eq: dS freq}
\omega_k^{\rm{dS}}(\eta)=\sqrt{k^2 +\mu^2/\eta^2},
\end{equation}
where we have defined  
\begin{equation}\label{muparam}
    \mu = \frac{1}{H_0}\sqrt{\frac{m^2 + 12(\xi-1/6)H_0^2 + 24\gamma H_0^2}{1 -12 \gamma H_0^2}}.
\end{equation} 
Note that $\mu$ behaves as a (constant) effective mass. Thus we see that the effect of the different considered couplings to the curvature is to modify the effective mass of the field during the inflationary phase.
Inspecting the possible values of the coupling constants, we note that:
\begin{enumerate}
    \item The effective mass $\mu$ becomes imaginary for certain combinations of $\xi$ and $\gamma $. In fact, the potential in these cases is not even bounded from below. These scenarios correspond to tachyonic states for the field.
    \item For $m=0,\ \xi=1/6,\ \gamma   = 0$ the effective mass vanishes and we recover the well-known result  that there is no production of purely conformal massless particles.
\end{enumerate}
Effectively, we have only a single parameter to characterise the scalar field, meaning that we cannot discriminate between different sets of the parameters $m$, $\xi$ and $\gamma$ leading to the same $\mu$ in pure de Sitter spacetime. 

The adiabatic condition \eqref{adiabatic} in this scenario is fulfilled either for large effective masses ($\mu \gg 1$) or whenever the combination $k|\eta| \gg 1$. From \eqref{muparam} we see that a positive value of the $\xi$ parameter tends to move the system towards the adiabatic regime, whereas nonvanishing values of the derivative coupling parameter $\gamma$ will move it away from this regime.

The solution to the mode equation \eqref{eqmodes} can be written in terms of   Hankel functions~\cite{AS}:
\begin{equation}
    \chi_{\mathbf{k}}(\eta) = \sqrt{k|\eta|}\big[A_{\mathbf{k}} H_\nu^{(1)}(k|\eta|) + B_{\mathbf{k}} H_\nu^{(2)}(k|\eta|)\big],
    \label{eq:hankel}
\end{equation}
where $\nu = \sqrt{1/4 -\mu^2}$.

The so-called Bunch-Davies vacuum is defined by mode functions that in the asymptotic past $\eta\to -\infty$ behave as plain waves \cite{ParkerToms} in analogy with the definition of vacuum in Minkowski spacetime. 
Demanding that in the asymptotic past the vacuum state only has positive-frequency modes, and taking into account the asymptotic behaviour of the Hankel functions for $|\eta|\to\infty$ \cite{AS}, the mode function $v_k$ must have the form \eqref{eq:hankel} with
\begin{equation}\label{eq: initial vacuum}
    A_{\mathbf{k}} = \sqrt{\frac{\pi}{2k}e^{i\pi\nu}}, \qquad B_{\mathbf{k}}=0.
\end{equation}

In order to simplify the computations and for later convenience, we will perform a change of variables from the set $(k,\eta)$ to the new set $(y,\eta)$, where the new variable is defined as $y = k|\eta|/|\nu|$. In this new set of variables, the modes defining the Bunch-Davies vacuum read: 
\begin{equation}
    \chi_y(\eta) = \sqrt{\frac{\pi|\eta|}{2}e^{i\pi\nu}}H^{(1)}_\nu(y|\nu|).
\end{equation}
It is now easy to see, after some algebra, that in terms of the independent variables $(y,\eta)$ the Bogoliubov coefficients $\beta_k(\eta):=\beta(y)$ depend only on the variable $y$ and not on $y$ and $\eta$ independently.

Once the $\beta(y)$ Bogoliubov coefficient is computed, the created number density of particles is obtained by integrating its modulus squared over all the possible momenta and dividing by the fiducial volume $a(\eta)^3$:
\begin{equation}\label{density}
     n(\eta) = \frac1{2\pi^2 a(\eta)^3}\int_0^\infty \, |\beta_k(\eta)|^2\, k^2 \mathrm{d} k = \frac{H_0^3|\nu|^3}{2\pi^2}\int_0^\infty \, |\beta(y)|^2\, y^2 \mathrm{d} y.
\end{equation}
Note that this expression is independent of the conformal time $\eta$. This is a remarkable result: the density of produced particles  in cosmological de Sitter spacetime is constant over time for a comoving observer in the adiabatic regime. 

\begin{figure}
    \centering
    \begin{overpic}[width=.45\textwidth]{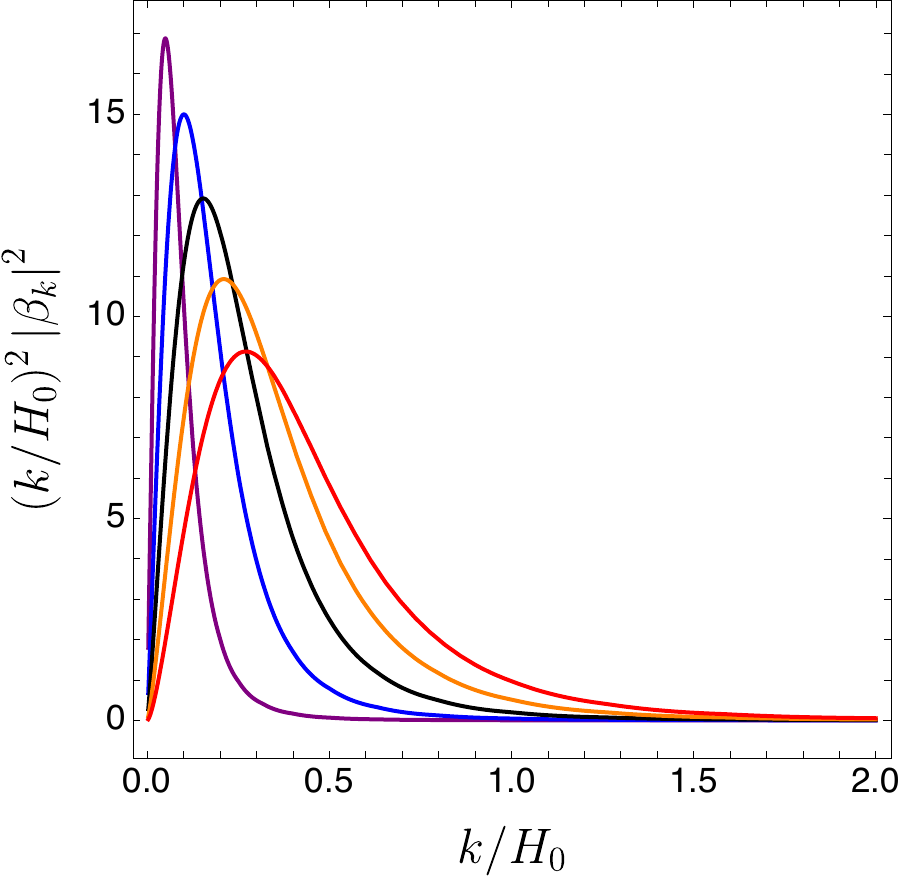}
    \put(14,101){{\scriptsize \sansmath $(\times 10^{-3})$}}
    \end{overpic}\qquad
    \begin{overpic}[width=.45\textwidth]{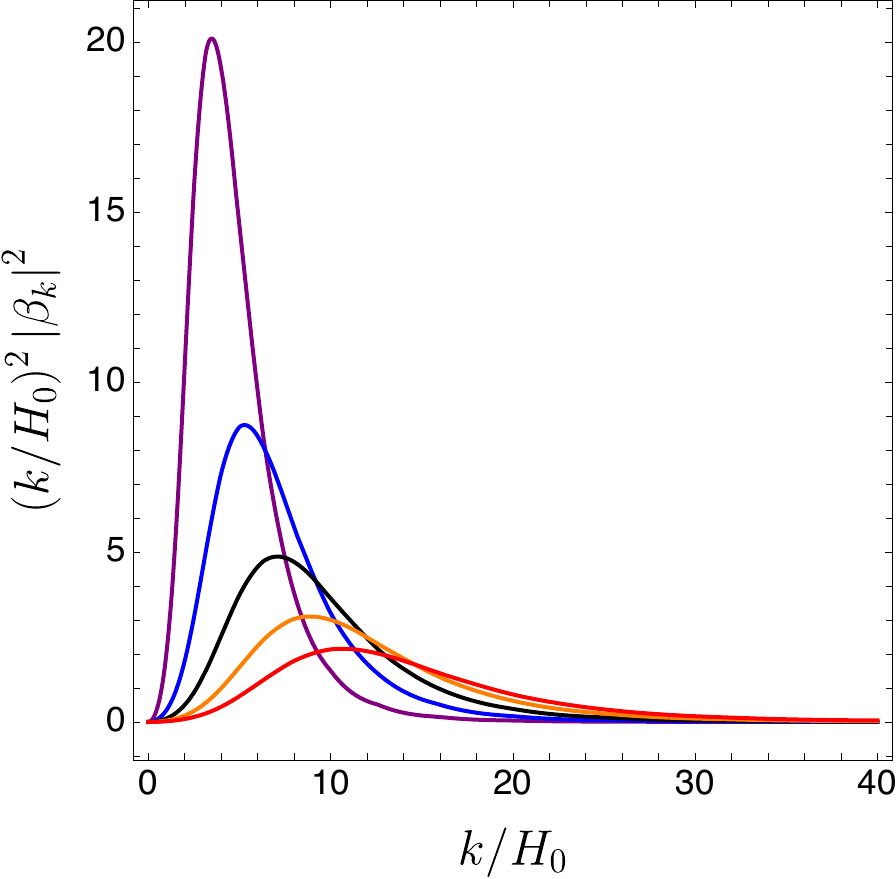}
    \put(14,101){{\scriptsize \sansmath $(\times 10^{-5})$}}
    \end{overpic}
    \caption{Spectral particle distribution created during the de Sitter epoch for different effective masses computed at $\eta=-1/H_0$. The left panel shows the regime of sub-Hubble masses with the following colour code: Purple, $m = 0.1 H_0$; Blue, $m = 0.2 H_0$; Black, $m = 0.3 H_0$; Orange, $m = 0.4 H_0$; and Red, $m = 0.5 H_0$. The right panel shows the regime for super-Hubble masses with the following colour code: Purple, $m = 4 H_0$; Blue, $m = 6 H_0$; Black, $m = 8 H_0$; Orange, $m = 10 H_0$; and Red, $m = 12 H_0$.}
    \label{fig:ds}
\end{figure}

In figure \ref{fig:ds}, we show the spectral particle production for small and large masses as compared with $H_0$ at the time $\eta=-1/H_0$. In the left panel we see    that as the mass increases within the sub-Hubble regime, the amplitude of the production peak decreases while the resulting spectrum  broadens, in such a way that the total production grows   as the mass increases. This result can be easily understood because the particle creation is a consequence of the conformal symmetry breaking, and the parameter controlling the degree of breaking is precisely the mass. On the other hand, for masses beyond the Hubble parameter $H_0$ (right panel), the peak amplitude also decreases as the mass increases but the spectrum broadens much less as compared with the left panel (not enough to overcome the dampening of the amplitude). The  global result is quite different: the net production decreases with increasing masses.  We can understand this effect if we think of the Hubble parameter as an indicator of the energy stored in the gravitational field. Then, creating particles   more massive than the energy of the background is more difficult.

We want to analyse the asymptotic behaviour of the number density \eqref{density} for the regime of large effective masses $\mu\gg 1$. In this regime, $\nu$ is pure imaginary, i.e. $\nu=i|\nu|$.
Hence, we will perform a series expansion of the mode solutions in $1/|\nu|$. One could be tempted to ignore  the contribution $1/4$ to  $\nu$ but it turns out to be crucial. Then, the asymptotic series for the Hankel function reads \cite{AS}
\begin{equation}\label{Happrox}
    H_{i|\nu|}^{(1)}(y|\nu|) = \sqrt{\frac{2}{i\pi|\nu|}} \frac{e^{\pi|\nu|/2} e^{i|\nu|\zeta}}{(1+y^2)^{1/4}} \left[1+i\frac{3p-5p^3}{24 |\nu|}
    -\frac{81 p^2 - 462 p^4 + 385 p^6}{1152 |\nu|^2} +\mathcal{O}\left(|\nu|^{-3}\right)\right],
\end{equation}
where
\begin{equation}
    \zeta(y) = \sqrt{1+y^2} + \log\left(\frac{y}{1+\sqrt{1+y^2}}\right),
    \qquad[p(y)=-(1+y^2)^{-1/2}].
\end{equation} 

When we compute the coefficient $\beta(y)$, the first and second terms of the series expansion cancel out, so the leading contribution to the particle density comes from the third order terms in \ref{Happrox}. Then, the leading order for $|\beta(y)|^2$ is
\begin{equation}
   |\beta(y)|^2 = \frac{(1+6 y^2)^2}{2^8 (1+y^2)^6 |\nu|^4} + \mathcal{O}(|\nu|^{-6}).
\end{equation}
It should be noted that for masses larger than a few times $H_0$ the agreement with the exact spectrum is almost perfect.

Finally, integrating \eqref{density} using the residue theorem, we find that the behaviour of the particle density in the large effective mass regime is
\begin{equation}\label{asympdensity}
    n(\eta) = \frac{151}{2^{18} \pi}\frac{H_0^3}{|\nu|}+\mathcal{O}(|\nu|^{-3}).
\end{equation}

\section{Particle creation during reheating}\label{multistagesec}

\subsection{Background evolution}

Our current understanding of the early universe includes two phases before the usual big bang cosmology: inflation and reheating.  The dynamics of both phases can be described as a FLRW model sourced by a scalar field which has a slow-roll regime, giving rise to the inflationary phase; it also has a minimum in its potential where it oscillates provoking its decay into the particles that permeate the universe nowadays in the process known as reheating. As we are interested in the effects of the derivative couplings discussed above in the gravitational production during the early universe, we need to take into account not only the inflationary phase, which is well described for our purposes with a de Sitter phase, but also we need to include the reheating epoch. We are mimicking the background spacetime dynamics of both phases using a de Sitter phase for inflation and a matter-dominated universe as the reheating phase. This choice of a matter-dominated solution comes from the mean behaviour of the background during the oscillations of the inflaton field $\phi$ in the massive chaotic inflationary scenario with a potential $\propto m_\phi^2\phi^2$ \cite{PhysRevD.56.3258}.

We can use the parameterisation for the scale factor introduced in \cite{PhysRevD.48.647} which provides a continuous and differentiable function:
\begin{align}\label{eq: scale factor}
	a(\eta)&=\begin{cases}
		\displaystyle \frac{1}{1-H_{0}\eta}, \qquad &\eta\leq0,\\
		\displaystyle\left(1+\frac{H_{0}\eta}{2}\right)^2, \quad &\eta>0,\end{cases}
\end{align}
where $H_{0} = \sqrt{ 12\pi} m_\phi$ is the value of the Hubble parameter at the onset of the single-field inflationary epoch for the model we are considering and we have set the end of inflation at $\eta=0$  (with $m_\phi $ being the inflaton mass). From now on    we will use $H_0$ as the basic quantity for units. This parameterisation is good enough to get a continuous Hubble parameter. However, it fails to provide continuous and differentiable curvature scalar and tensor. In order to have a satisfactory   model, we need the frequency of our field to be sufficiently smooth.  Hence, we need to introduce an interpolation regime for the curvature quantities between the end of inflation (i.e., the end of the de Sitter phase) and a moment $\eta_*$ in the reheating epoch. This time $\eta_*$ is chosen by equating the Hubble parameter from the above parameterisation and the one obtained through the dynamics of the inflaton field, which for any single-inflation model can be written during the oscillatory phase of the field around the minimum of its potential as
\begin{equation}\label{eq: Hubble parameter}
	H^2 = \frac{8 \pi}{6 M_{\rm{P}}^2}\left(\frac{1}{a^2} {\phi'}^2+m_\phi^2\phi^2\right), \quad \phi = \frac{\Phi_0}{\eta + \frac{H_0\eta^2}{2}+\frac{H_0^2\eta^3}{12}}\sin\left[m_\phi\left(\eta + \frac{H_0\eta^2}{2}+\frac{H_0^2\eta^3}{12}\right)\right],
\end{equation} 
where the inflaton field has been expanded at first consistent WKB order. In these expressions, \mbox{$M_{\rm{P}}=1.22\cdot 10^{19}{\rm{GeV}}$} is the Planck mass, $m_\phi= 10^{13} \rm{GeV}$ is the inflaton mass,  and $\Phi_0 = M_{\rm{P}}/(2 \sqrt{\pi})$ is the amplitude of the field at the onset of the oscillations. These values for the inflaton mass and amplitude are constrained by the CMB observations \cite{Ade:2015xua}. Equating \eqref{eq: Hubble parameter} and the Hubble parameter from the parameterisation of the scalar factor, we determine $\eta_*=2.4/H_0$.

\subsection{Particle production}

In a model driven by an inflaton field, all the background quantities would be sufficiently smooth functions. Hence, we want to interpolate $R$ and $R_{\mu\nu}$ so they reproduce this differentiability to sufficiently high order. In order to do so, we introduce  polynomials of degree 24 for the value of the functions as well as for each of their derivatives independently, so each of them is a $\mathcal{C}^{12}$ function. The degree of the polynomials has been chosen so it is the minimum possible one so the results are independent of the interpolation order. Note that without the interpolation, the frequency would exhibit a discontinuity when the de Sitter epoch finishes. This discontinuity would abruptly excite modes of arbitrarily high momenta on the field and hence the expected production of particles would be greater than the one obtained considering the background evolution due to an inflaton field.

The equation of motion for the modes during the post inflationary epoch, after the junction time $\eta_*$, acquires the form
\begin{equation}\label{multistage}
    \chi_k'' + \omega^{\rm{pi}}_k(\eta)^2\chi_k = 0,
\end{equation}
where the frequency of the modes has a rather complicated functional form in terms of the background quantities:
\begin{equation}
	\omega_k^2 = \frac{1+B}{1+A}k^2+F,
\end{equation}
where $A,\,B,$ and $F$ are given in \eqref{functions} particularised for the background described above.

The expression of the frequency as a function of the conformal time $\eta$ is obtained using the Hubble parameter, the Ricci tensor,  and the curvature scalar coming from the averaged behaviour of the inflaton field during the reheating epoch. We will solve the equation for the modes of the field numerically, using as initial conditions the value of the field and its first derivative at the end of the de Sitter phase, given by \eqref{eq:hankel} with the constants given by \eqref{eq: initial vacuum}. Despite the cumbersome expression for the frequency, a direct evaluation of the adiabatic parameter $|\omega_k'/\omega_k^2|$ yields that we safely lie in an adiabatic regime on the reheating phase for all the region of the parameter space we will consider. Therefore the definition of the adiabatic vacuum is well suited to define a proper instantaneous vacuum to compare with the evolved initial state.

Now we are ready to compute the gravitational particle production during the early universe using the model we have discussed, delving into the influence of the derivative couplings in this process. Once we have computed the numerical mode functions which define  the initial vacuum state, we can compute the Bogoliubov coefficients relating this initial vacuum state and the adiabatic ones during the post inflationary epoch via equation~\eqref{coefs}. 

We will consider the derivative couplings as perturbations, meaning that their contribution is subdominant with respect to the one coming from the mass term and the coupling to the Ricci scalar. Furthermore, even for perturbative values of these derivative couplings there is a strong limitation coming from the de Sitter phase, as each case in which their contributions lead to imaginary effective masses, i.e., $\mu^2<0$, the vacuum state of the field would become unstable \cite{BirrelDavies}. Hence, we have an additional constraint on the derivative couplings given by the inequality:
\begin{equation}\label{eq:limits}
     - 2[m^2/{H_0^2}+12(\xi-1/6) ]\leq 12\gamma H_0^2 < 1.
\end{equation}

\subsection{Perturbative analysis} \label{sub:perturbative}

For sufficiently small values of the derivative couplings $\gamma$ and $\sigma$, we can perform an expansion of the particle production.
As we argue in the following this requires two different regimes of approximation: a perturbative expansion of the action that can still induce nonlinear effects due to the nontrivial kinetic term, and an additional expansion for small derivative couplings as compared with the mass (in appropriate units). In this subsection, we will highlight the main aspects of these perturbative and linear expansions for the particle production.

There are reasons that suggest that we should restrict the values of the derivative couplings to the perturbative regime. For instance, these derivative couplings affect the kinetic term of the field, modifying the propagator in a highly nontrivial way. In particular, the derivative couplings of the scalar field to the Riemann tensor at first order are not renormalisable. If such terms are present, higher order terms are expected. This means that if the first order contributions are not small, neglecting higher terms may not be well-motivated.

In view of the action \eqref{eq: action}, we can qualitatively advance that, in the perturbative regime, the coupling $\sigma$ multiplied by the largest value of the traceless Ricci tensor along the whole period relevant to particle production must be much smaller than 1. Similarly, $\gamma$ multiplied by the largest value of the curvature scalar must be much smaller than 1.  More explicitly, it is easy to see that for metrics of the form \eqref{eq:flrw} the factor multiplying to  ${\varphi'}^2$ is proportional to $1+A$ and the factor multiplying to $\partial_i\varphi\partial^i\varphi$ is proportional to $1+B$, where $A$ and $B$ are given by in \eqref{functions}. The perturbative regime is then achieved when $|A|$ and $|B|$ are much smaller than 1 during the whole evolution. Considering the maximum values of the background quantities multiplying $\gamma$ and $\sigma$, we can write the perturbative conditions in terms of conditions on the maximum absolute value of the derivative couplings:
 \begin{equation}
      |\bar\gamma| \ll 1,\qquad   |\bar\sigma|\ll 15.
     \label{eq:pertublim}
\end{equation} 
Here and from now on we will use the rescaled dimensionless derivative couplings \mbox{$\bar\gamma=12 H_0^2\gamma$} and $\bar\sigma= H_0^2\sigma$, and the dimensionless mass parameter $\bar m=m/H_0$.

Even in this perturbative regime, the way in which the quantity $A$ enters  the definition of the quantisation field $\chi$ makes the final result nonlinear in the parameters $\bar \gamma$ and $\bar\sigma$. In particular, the derivative coupling $\bar\gamma$ contributes to the effective mass of the field, even during the de Sitter phase, while $\bar\sigma$ does not (see e.g. Eq. \eqref{muparam}). In the following we focus on the case $\xi=1/6$ both for simplicity and to isolate the effects of the derivative couplings. 

Let us check the conditions on the derivative couplings that are necessary for the  changes on the particle production to be small. To begin with,  we expand the  frequency to first order in the derivative couplings,
\begin{align} 
    \omega_k^2 &=\omega_{0k} ^2 + \Omega_k ,
    \nonumber\\
       \omega_{0k}^2 &= k^2+ a^2 m^2 ,
    \nonumber\\
     \Omega_k &=(B-A)k^2+ a^2(\bar m^2 H_0^2+R/6) A -a H A'-A''/2.
\end{align}
If we impose the condition that the linear term $\Omega_k$ is small compared with the unperturbed frequency squared $\omega_{0k}^2$, once we evaluate the background geometrical quantities accompanying the derivative coupling at their maximum value during the whole evolution, we can extract a necessary condition on the derivative parameters:
 \begin{equation}
     |\bar\gamma| \ll     \bar m^2 ,\qquad     
     |\bar\sigma|\ll \frac{15 \bar m^2 }{2+\bar m^2},
     \label{eq:pertublim2}
\end{equation}
These conditions are necessary but not sufficient to have small  changes in the particle production~$|\beta_k|^2$ as we discuss in the following.
  
Let us start  will the zeroth order $|\beta_{0k}|^2$. This is calculated using the zeroth order term $u_{0k}$ of  the adiabatic modes $u_k$ (i.e., \eqref{adiabaticos} with $\omega_{0k}$ instead of $\omega_k$) and the zeroth order $v_{0k}$ of the vacuum modes $v_k$.
These zeroth order modes are solutions to the equation $\ddot v_{0k}+\omega_{0k}^2 v_{0k}=0$. 

The contribution to $|\beta_k|^2$ linear in the derivative couplings can also be easily found, although the calculations are quite tedious and not very illuminating. Let us describe the ingredients involved in the final expression other than the zeroth order quantities already mentioned, namely, $\omega_{0k}$, $u_{0k}$, $v_{0k}$, and their derivatives. These extra ingredients are the  already calculated  linear order contribution to the frequency $\Omega_k$ and its derivatives, the corresponding perturbations to the adiabatic modes, and the perturbation $\zeta_k$ to the modes~$v_k$, defined as $v_k=v_{0k}(1+\zeta_k)$.
This is the only nontrivial quantity that we have to calculate. Taking into account the equation  that it  must satisfy (namely, \eqref{eqmodes}   with $\omega_k$ expanded to first order), it is easy to see that $\zeta_k$ is given by the quadrature
\begin{equation}\label{eq: mode pert}
    \zeta_k(\eta)=-  \int_{-\infty}^\eta d\eta'\frac1{v_{0k}(\eta')^2}\int_{-\infty}^{\eta'} d\eta'' v_{0k}(\eta'')^2\Omega_k(\eta'').
\end{equation}
In view of this expression, we can assume that $\zeta_k(\eta)$ is small throughout the evolution provided that $\bar \gamma$ and $\bar \sigma$ satisfy some extra conditions. 

For illustrative purposes let us study the large mass regime.
Taking into account the form of $\Omega_k$ and a quick numerical estimation of the values of the time integrals involved, we obtain the following rough estimation of the maximum value throughout the evolution for $\zeta_k(\eta)$:
\begin{equation}
|\zeta_k(\eta)|\lesssim   \frac{\bar m}{12}\big(|\bar\gamma| +
    \bar m^{-7/2} |\bar\sigma|\big),
    \label{eq:zetalargemass}
    \end{equation}
which apply for large masses. The mass-dependent factors   come from the double time integral. The huge difference in their values can be traced back to the fact that  the dominant evolution quantity accompanying $\bar\gamma$ is proportional to $H(\eta)^2 \bar m^2$ and the  one accompanying $\bar\sigma $ is proportional to $H'(\eta) \bar m^2$: the former has a nonvanishing value during the de Sitter phase  and   decays afterwards while the latter vanishes in the de Sitter phase and grows in absolute value from there to vanish again. The  double integration with the highly oscillatory function $v_{0k}(\eta)^2$ (with a mass dependent frequency) then reduces the amplitude of the $\bar\gamma$ term by a factor $1/\bar m $, while that of $\bar\sigma$  has as an additional  factor $1/\bar m^{7/2}$. The linear modification of $|\beta_k|^2$ also depends on the derivative of the mode but nothing new comes from there.
We learn two things from this upper  bound to $\zeta_k$. First, the contribution of the term linear in $\bar \gamma$  is small in the perturbative regime.  Second, in the same regime, the contribution  of $\bar\sigma$ is negligible.
It turns out that, for large masses, although the linear contributions are small in the perturbative regime as we have argued, this does not mean that the quadratic terms in the derivative couplings are negligible  because the mass-dependent factors accompanying them can be very large and in fact dominate. As we will see in the next section, this is indeed the case for  the $\bar \gamma$ contribution. On the other hand, we will also see that the quadratic terms in $\bar\sigma$ are small in the large mass regime, provided that \eqref{eq:pertublim2} are satisfied, becoming absolutely negligible. 

For small masses, the linear analysis is less conclusive. However it seems to indicate that small values of $\sigma$ can lead to significant changes in the production.

Performing the second order expansion in order to check the conditions for the linear regime to be valid is calculationally very expensive, more so  taking into account that we can provide   numerical results to all orders and directly see the role of each of the couplings to the particle production. In fact, as already suggested above, the most interesting results are obtained within the perturbative regime but for  nonlinear contributions to the production. This is done in what follows.

\subsection{Results}

We will start our analysis discussing the effects of the different couplings on the spectral particle distribution for the field.  We will cover the perturbative   range  for the derivative couplings limited by the restriction \eqref{eq:limits} that ensures that there are no tachyons in the theory but allowing for nontrivial behaviours of the particle production.  

\subsubsection{Spectral particle production}

\begin{figure}
    \centering
    \begin{overpic}[width=.45\textwidth]{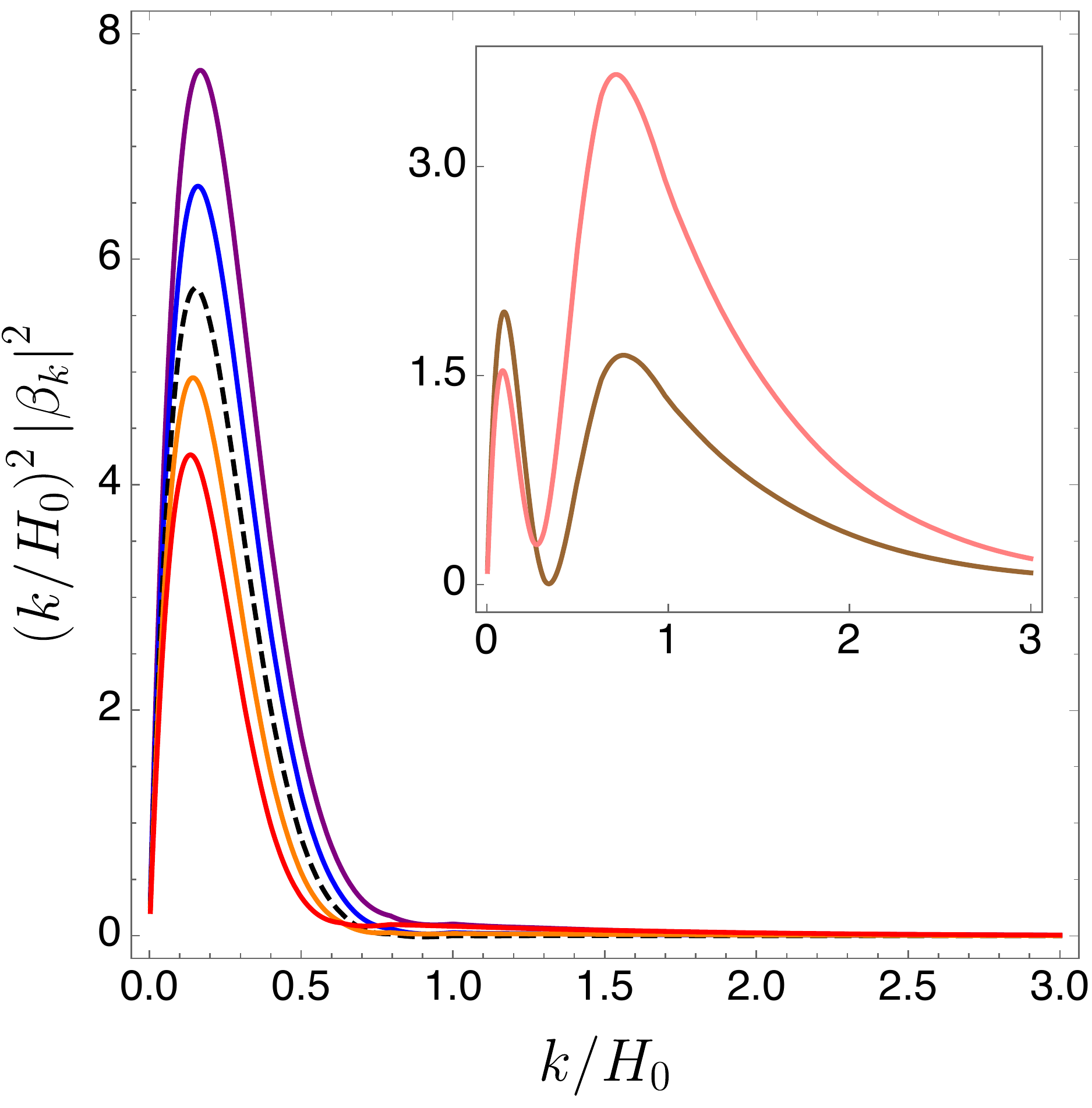}
    \put(12,102){{\scriptsize \sansmath $(\times 10^{-3})$}}
    \end{overpic}\qquad
    \begin{overpic}[width=.45\textwidth]{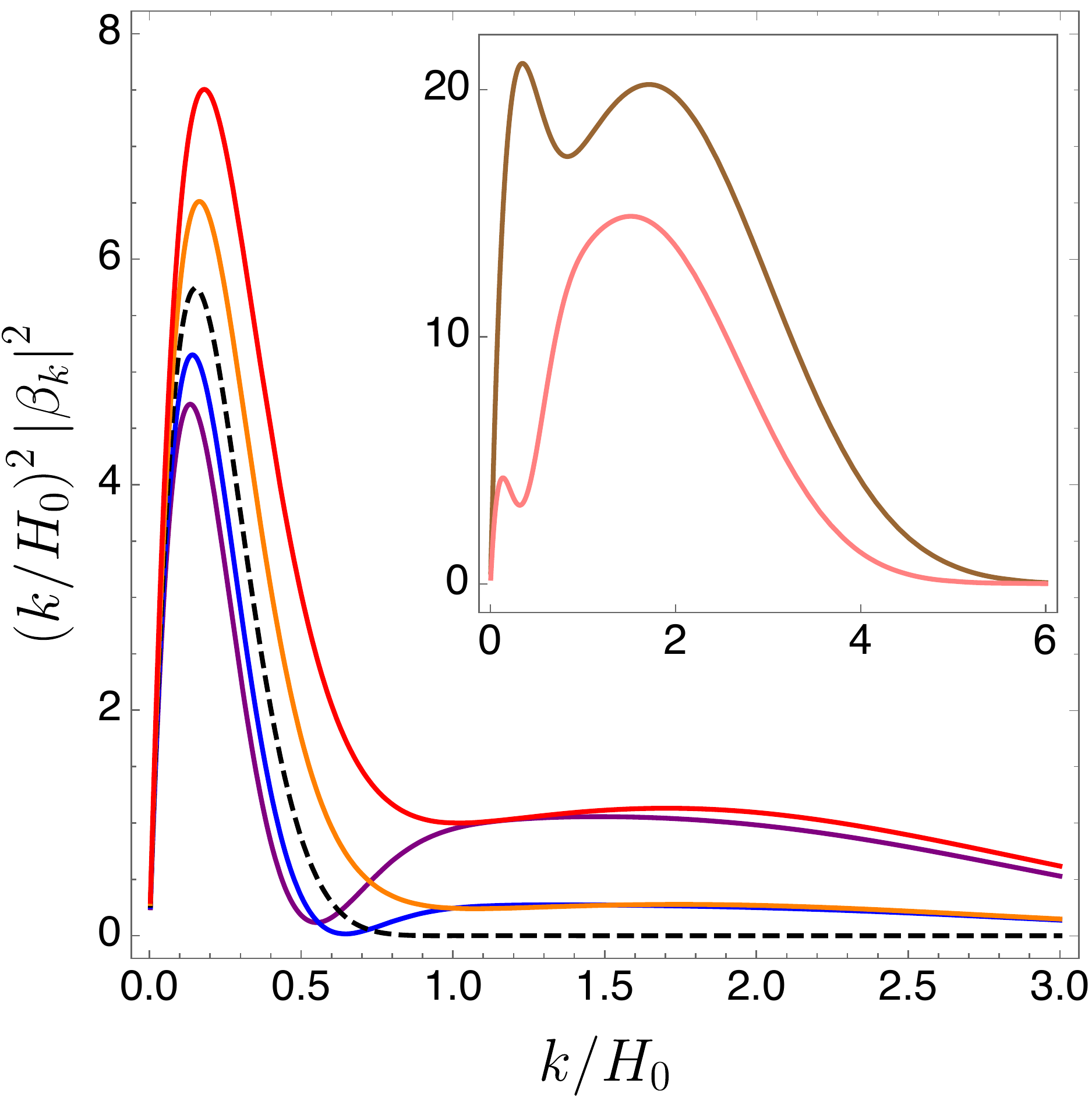}
    \put(12,102){{\scriptsize \sansmath $(\times 10^{-3})$}}
    \end{overpic}
    \caption{Spectral particle density for the case $\xi=1/6$ and $\bar m =0.1$ as a representative case for small masses at the end of reheating.  On the left panel, different curves correspond to different values of $\bar\gamma$ keeping $\bar\sigma=0$ with the following colour code: Purple, $\bar\gamma =-\bar m^2/2$;  Blue, $\bar\gamma=-\bar m^2/4$; Black (dashed), $\bar\gamma=0$; Orange, $ \bar \gamma =\bar m^2/4$; Red, $\bar\gamma= \bar m^2/2 $;   Brown, $\bar\gamma= 2 \bar m^2$; and Pink, $\bar\gamma= 3 \bar m^2$. The companion curves corresponding to $\bar\gamma=-2\bar m^2,-3\bar m^2$ are not shown because these values of $\bar\gamma$ are beyond the tachyonic limit \eqref{eq:limits}.
    On the right panel different curves correspond to different values of $\bar\sigma$ keeping $\bar\gamma=0$ with the following colour code: Purple,
    $\bar\sigma =-1/2$;  Blue, $\bar\sigma=-1/4$; Black (dashed), $\bar\sigma=0$; Orange, $ \bar \sigma =1/4 $; Red, $\bar\sigma= 1/2$;  Brown, $\bar\sigma= 2$; and Pink, $\bar\sigma= -2$.}
    \label{fig:spectrasmallmass}
\end{figure}

Figure \ref{fig:spectrasmallmass} shows the spectral particle production for small masses (we use $\bar m=0.1$ as the working value). The left panel shows the production for various values of the derivative coupling $\bar \gamma$ keeping $\bar \sigma=0$, while the right panel shows the production  for various values of $\bar \sigma$ keeping $\bar \gamma=0$, so we can study the effect of each coupling separately. Both panels are very different as far as the considered range of the derivative couplings is concerned. In the left panel, the main graphics correspond to values of the $\bar\gamma$ coupling (not much) smaller than the  boundaries \eqref{eq:pertublim2} of the regime in which we expect small variations in the production. The inset displays the production for couplings beyond this regime but still well inside the perturbative limit \eqref{eq:pertublim}. In the right panel, however, the main graphics correspond to values of the $\bar\sigma$ coupling  close to 1.
These values are inside both the perturbative regime and the range of \eqref{eq:pertublim2}. The inset displays the production for couplings also inside this regime but with significant nonlinear changes.
This choice of values for the derivative coupling $\bar\sigma$ (associated with the traceless Ricci tensor) is motivated by the fact that $\bar\sigma$ has a much smaller effect on the particle production (as compared with $\bar \gamma$). This is suggested by the arguments and bounds discussed in the previous subsection and confirmed by the numerical results.
 
Indeed, for values of $\bar\sigma$ comparable to those of $\bar\gamma$ in the left panel, the result would be just that all curves are superimposed. Even more, as we can see in figure \ref{fig:spectrasmallmass} the change in the spectral production for   values of $\bar\sigma$ close to 1 are of the same order as for values of $\bar\gamma$ inside the regime \eqref{eq:pertublim2}.
The specific ranges that we have used for these couplings lie in the intervals $\bar\sigma \in [-1/2,1/2]$ in the main graphics. For the inset we have considered the values $\bar\sigma = \pm2$, still inside both the perturbative limit and the regime \eqref{eq:pertublim2}, but for which we obtain nontrivial contributions. We have considered $\bar\gamma\in[-\bar m^2/2,\bar m^2/2]\subset(-1,1)$ (with the regime in which we expect no significant changes being $|\bar\gamma|\ll \bar m^2 $ as deduced from \eqref{eq:pertublim2}). Note that we cannot go beyond $\bar\gamma=-\bar m^2$ due to the restriction~\eqref{eq:limits} even being within the perturbative regime. For the curves on the inset we have considered the values $\bar\gamma=(2\bar m^2,3\bar m^2)$ for which we expect nontrivial contributions to particle production, though still well inside the perturbative limits.

Let us start with the main graphics on the left panel (varying $\bar \gamma$). It is not surprising that within the considered range for $\bar\gamma$ the changes in the amplitude of the spectra are linear in the value of the coupling. We see that for any value of $\bar\gamma$ the spectral production has the same dominant peak as in the conformal case and that its amplitude grows for negative values of $\bar\gamma$ and diminishes for positive values.  
There are two a priori factors behind these changes in the peak amplitude. First, the curvature scalar (coupled through $\bar\gamma$) contributes to the value of the effective mass of the field during the de Sitter phase diminishing and broadening the spectral production for larger effective masses (see left panel in figure \ref{fig:ds}). Second, $\bar \gamma$ also contributes through an additional pure curvature effect, present only after the de Sitter phase, when the curvature scalar evolves with time. 
Let us briefly discuss this mechanism in more detail. If we consider the evolution after the de Sitter phase with only a mass term equivalent to the effective mass (including the effect of $\bar \gamma$ on the mass, i.e., $\mu(\bar m,\bar\gamma)$, but neglecting any other effect during the evolution) we see that the behaviour of the amplitude of the spectra with $\bar \gamma$ is inverted with respect to that in figure \ref{fig:ds}, with all the peaks centred around the same momentum and with similar width. This is precisely the opposite behaviour to that in the left panel of figure \ref{fig:spectrasmallmass}. 
For instance, note that the purple curve (corresponding to $\bar\gamma=-\bar m^2/2$) gives an effective conformal behaviour ($\mu=0$) in the de Sitter epoch, hence no particle production occurs during that epoch, and still the spectral production is significantly augmented during the transition phase.  

The inset in the left panel shows the spectra for situations beyond the regime \eqref{eq:pertublim2} but still for perturbative values of the coupling $\bar\gamma$.  We can see that the effect of the coupling to the curvature scalar induces significant changes to the spectral production as a secondary peak emerges at higher momenta. This secondary peak dominates the production for values of the coupling beyond the linear regime. 

The main graphics on the right panel (varying $\bar\sigma $ close to 1 and with $\bar \gamma=0$) also show a dominant production peak for momenta that were sub-Hubble during inflation. In this case, there is no contribution of the derivative couplings to the effective mass $\mu$ and the whole effect is due to pure tensor (i.e. traceless) Ricci curvature enhancements. We also see that the production for large momenta is significantly enhanced by the very same mechanism over a broad region of the spectra, augmenting the production even further. This effect is specially important on the red and purple curves which correspond to the larger values of $\bar\sigma$ within the considered range. In the inset we can see that the spectral production grows wildly, despite $\bar\sigma$ being inside the regime \eqref{eq:pertublim2}, as discussed in the previous subsection.  For positive values of $\bar\sigma$ there are two contributions: the primary peak corresponding to the de Sitter production enhanced by the subsequent evolution and a broad band on larger momenta, while for the negative values all the production is concentrated on the latter broad band.

As a final remark, we can conclude that the coupling to the traceless Ricci tensor contributes significantly less than the coupling to the curvature scalar as we have already inferred from the perturbative analysis of the contributions to the modes.

\begin{figure}
    \centering
    \begin{overpic}[width=.46\textwidth]{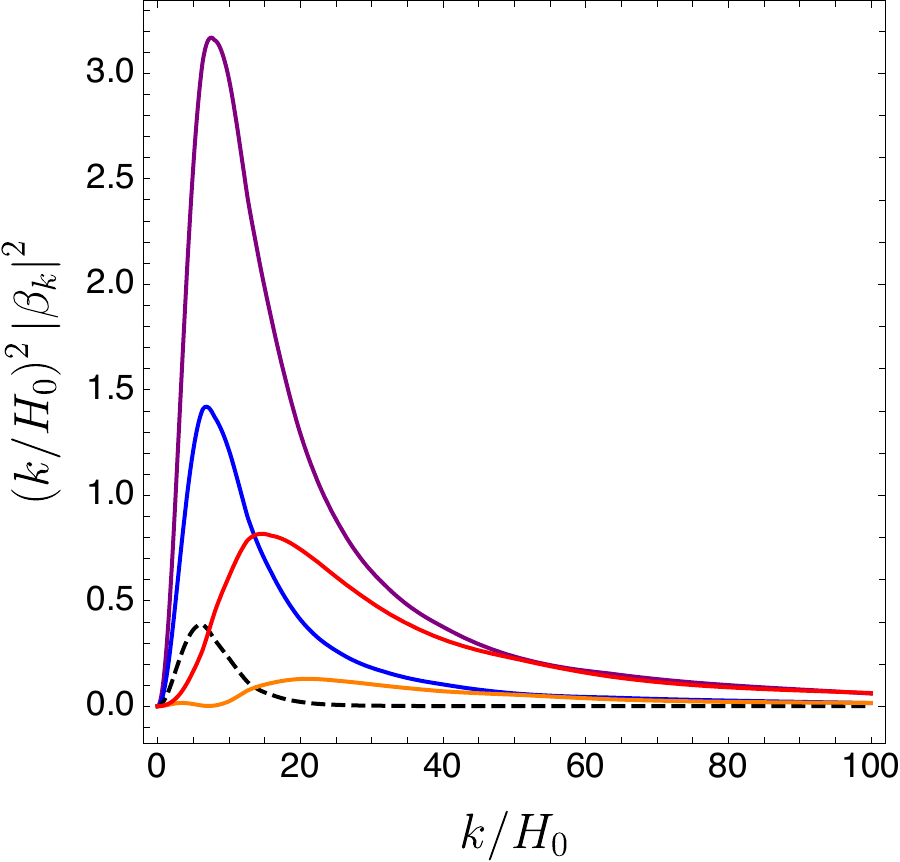}
    \put(15,99.5){{\scriptsize \sansmath $(\times 10^{-4})$}}
    \end{overpic}\qquad
    \begin{overpic}[width=.445\textwidth]{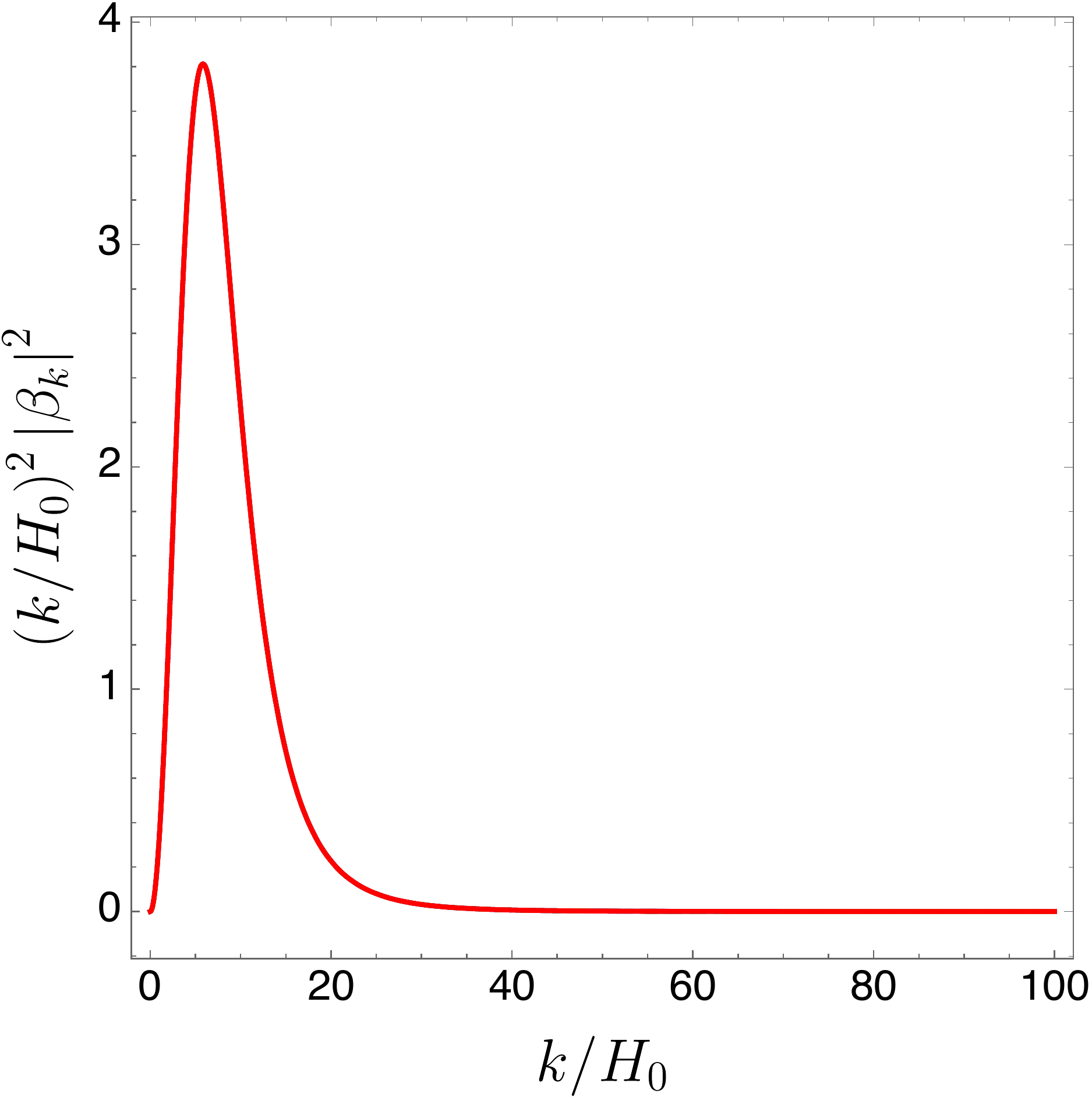}
    \put(12,102){{\scriptsize \sansmath $(\times 10^{-5})$}}
    \end{overpic}
    \caption{Spectral particle density for the case $\xi=1/6$  and $\bar m=10$ as a representative case for large masses at the end of reheating. On the left panel, different curves correspond to different values of $\bar\gamma$ keeping $\bar\sigma=0$ with the following colour code: Purple, $\bar\gamma =-0.01/2$;  Blue, $\bar\gamma=-0.01/4$; Black (dashed), $\bar\gamma=0$; Orange, $ \bar \gamma =0.01/4$; Red, $\bar\gamma= 0.01/2 $. On the right panel, different curves correspond to different values of $\bar\sigma$ keeping $\bar\gamma=0$ with the following colour code: Purple, $\bar\sigma =-1/2$;  Blue, $\bar\sigma=-1/4$; Black (dashed), $\bar\sigma=0$; Orange, $ \bar \sigma =1/4$; Red, $\bar\sigma= 1/2 $.}
    \label{fig:spectralargemass}
\end{figure}

In figure \ref{fig:spectralargemass} we show the effects of the derivative couplings on the spectral particle production for large masses of the field (we have considered $\bar m=10$ as a representative value). The left panel shows the production for various values of the derivative coupling~$\bar \gamma$ for $\bar \sigma=0$, while the right panel shows the production for different values of $\bar \sigma$ keeping $\bar\gamma=0$. In both cases, the main graphics correspond to values of the considered derivative coupling smaller than the perturbative boundaries \eqref{eq:pertublim}. In particular, the parameters lie in the interval $\bar\gamma \in [-0.01/2,0.01/2]$, and $\bar\sigma\in[-1/2,1/2]$, respectively (the same as in figure \ref{fig:spectrasmallmass}). Note that these values are inside the regime \eqref{eq:pertublim2},   where we expect small linear effects  as discussed in subsection \ref{sub:perturbative}. 

The figure on the left panel depicts the dependence of the spectral particle distribution on different values of the derivative coupling $\bar\gamma$, within the perturbative regime, for $\bar\sigma=0$. 
We see that for negative values of the coupling constant $\bar\gamma$, the contribution of the curvature scalar is to enhance the particle distribution obtained at the end of the de Sitter phase, as the spectral production is dominated by the same single peak. On the other hand, for positive values of the coupling $\bar\gamma$ to the curvature scalar we see the same effect as for couplings beyond the linear regime for small masses (inset of figure \ref{fig:spectrasmallmass}-left), meaning that the de Sitter peak is damped and that there appears a broad band of production on larger momenta, becoming the dominant contribution to the spectral production. 
As happened for small masses, there are two different factors to explain these contributions: the changes in the de Sitter effective mass and the pure curvature effects during the transition phase. It is not difficult to see, by direct comparison with the right panel of figure \ref{fig:ds}, that the wild variation on the amplitude of the peaks has its major contribution from the effect of the curvature variation during the transition between inflation and reheating. It is important to note that even though we are considering values of the coupling constant well inside the perturbative regime and much smaller than the mass of the field, the resulting contributions to the particle distribution are nonlinear in the strength of the coupling $\bar\gamma$, as advanced in  subsection \ref{sub:perturbative}.

On the right panel, the main figure shows that the effect of the traceless Ricci tensor mediated by the $\bar\sigma$ coupling on the spectral production is negligible for values of the coupling within the perturbative limit inasmuch as all the curves lie one on top of the other. This is precisely the behaviour expected from \eqref{eq:zetalargemass}, where  the $\bar\sigma$  linear  contribution to the mode perturbation is very much suppressed by high powers of the mass.

Comparing the results for large masses with our prior analysis of the small mass regime we can see two important differences: on the one hand, the importance of the derivative coupling $\bar\sigma$ diminishes, going from giving important contributions to the spectral density (figure \ref{fig:spectrasmallmass}-right) to becoming negligible in the large mass scenario (figure \ref{fig:spectralargemass}-right). On the other hand, we see that for the same values of the coupling $\bar\gamma$ to the  curvature scalar, the contributions of this term go from being linear for small masses to become nonlinear in the large mass regime. Hence, we can conclude from the analysis of the spectral distribution that, for large masses, the produced density becomes very sensitive to the value of $\bar\gamma$, while it is not affected by changes in $\bar\sigma$.

\subsubsection{Total density of produced particles}

Once we have discussed the effect of the derivative couplings on the spectral distribution of produced particles, we are ready to focus on the total produced density at a time $\eta>\eta_\star $ after reheating
\begin{equation}\label{density2}
     n(\eta ) = \frac{H_0^3}{2\pi^2 a(\eta )^3} \bar N(\eta_\star ),
     \qquad
     \bar N(\eta )= \frac{1}{H_0^3}
     \int_0^\infty \, |\beta_k(\eta )|^2\, k^2 \mathrm{d} k .
\end{equation}
We take $\eta_\star=50/H_0$ as a working value. The subsequent evolution of $\bar N(\eta)$ is constant for $\eta>\eta_\star$. Therefore, the evolution of the density $n(\eta)$ will be solely due to the expansion of the background as the gravitational production is already stabilised. 

\begin{figure}
    \centering
    \begin{overpic}[width=.45\textwidth]{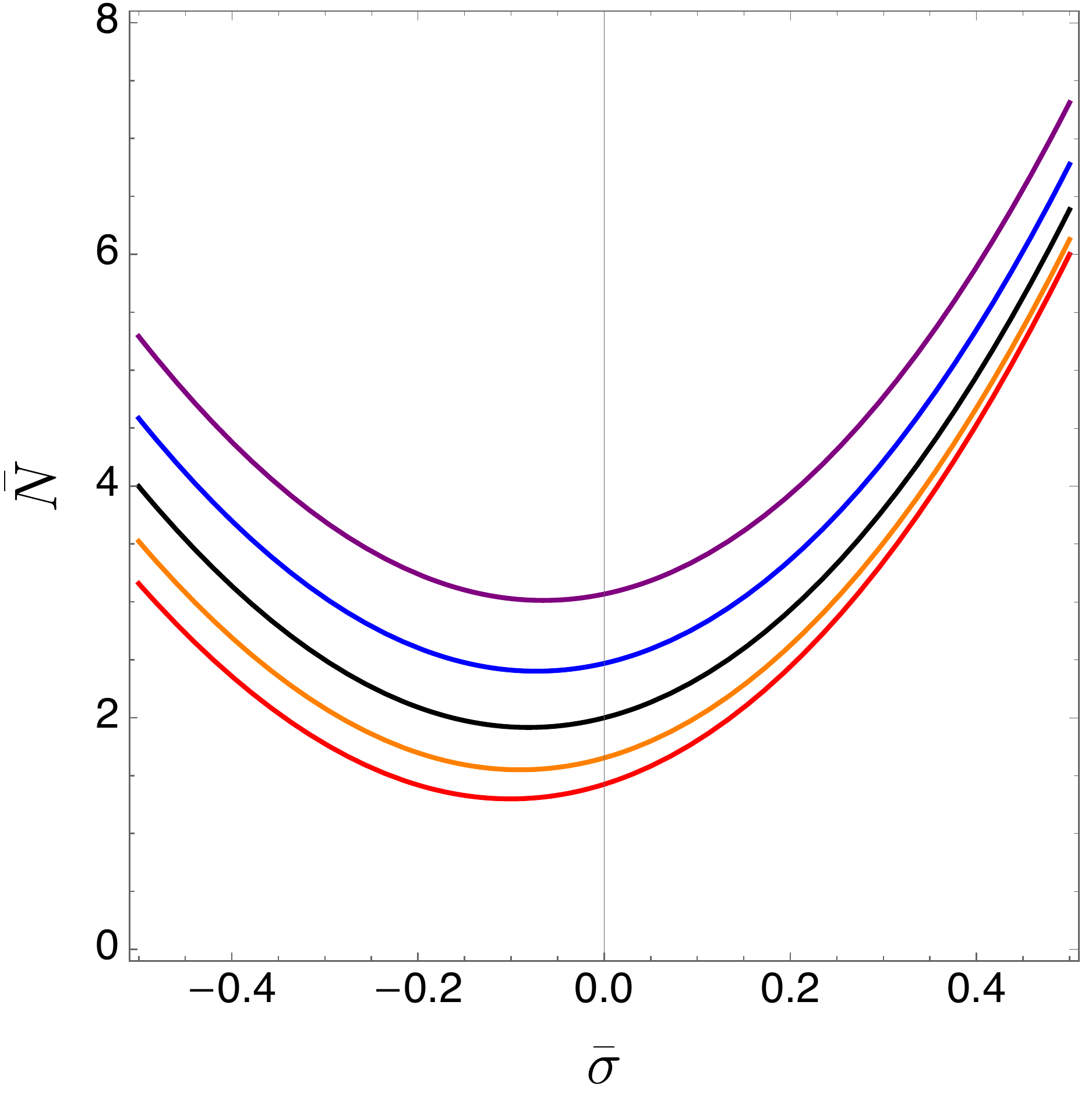}
   \put(11.5,102){{\scriptsize \sansmath $(\times 10^{-3})$}}
    \end{overpic}
    \qquad \begin{overpic}[width=.45\textwidth]{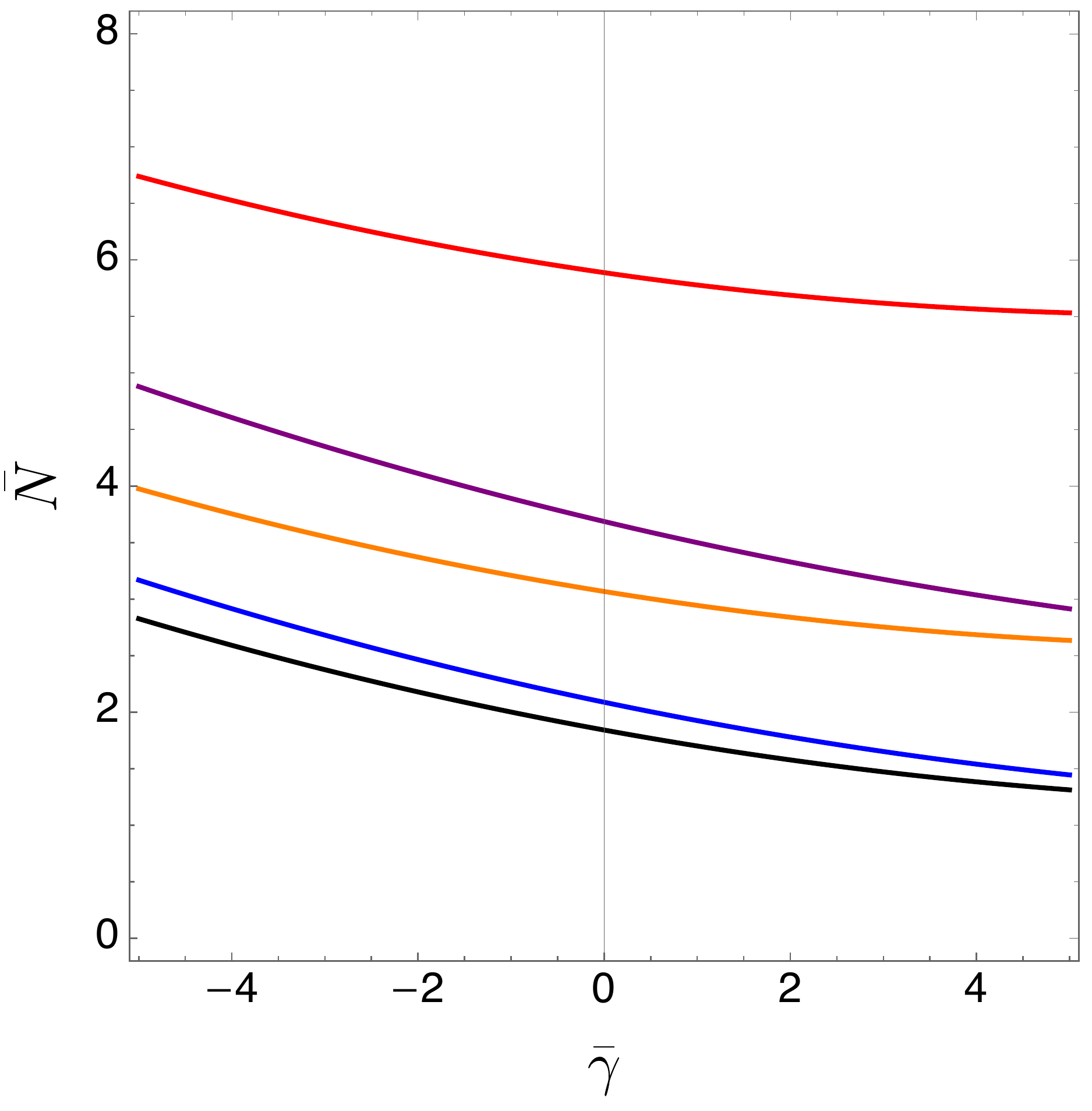}
    \put(11.5,102){{\scriptsize \sansmath $(\times 10^{-3})$}}
    \put(92,9){{\scriptsize \sansmath $(\times 10^{-3})$}}
    \end{overpic}
    \caption{Density of produced particles for the small mass ($\bar m=0.1$, $\xi=1/6$) scenario computed at $\eta_\star=50/H_0$.  On the left panel, we show  the density as a function of $\bar\sigma$ for different values of $\bar\gamma$ in the perturbative regime.
    The colour code is the following: Purple, $\bar\gamma = -\bar m^2/2$;  Blue, $\bar\gamma = -\bar m^2/4$; Black, $\bar\gamma = 0$; Orange, $\bar\gamma = \bar m^2/4$; Red, $\bar\gamma = \bar m^2/2$.
    The right panel  shows the density as a function of $\bar\gamma$ for different values of $\bar\sigma$ (in the perturbative regime) with the following colour code: Purple, $\bar\sigma = -1/2$;  Blue, $\bar\sigma = - 1/4$; Black, $\bar\sigma = 0$; Orange, $\bar\sigma =  1/4$; Red, $\bar\sigma =   1/2$.}
    \label{fig:smallmass}
\end{figure}

Let us start discussing the effects of the derivative couplings in the small mass regime for the field. 
In figure \ref{fig:smallmass} we show the number density of particles computed numerically at $\eta_\star$ for the small mass case ($\bar m=0.1$, $\xi=1/6$) as a function of $\bar\sigma$ and $\bar\gamma$. 

The graphics in the left panel depict the density of produced particles as a function of the derivative coupling  $\bar\sigma$ for several different constant values of $\bar\gamma$. As we can see and  have already mentioned in the previous analysis, we are in the nonlinear regime of this parameter except for values around $|{\bar\sigma}|\simeq 0.1$.  One of the main features of the dependence with $\bar\sigma$ is that the density attains a minimum value for some negative value of the coupling (actually the specific value varies with $\bar\gamma$) and grows significantly for positive and larger negative values, which is not surprising as we are already near the  limit \eqref{eq:pertublim2}. Although these curves may seem parabolic, they actually have higher order contributions.
Another important observation at the light of this figure is that the same density is produced for different pairs $(\bar\sigma, \bar\gamma)$. In particular, if we take the production in the case with no derivative couplings, we can see that the same total production can be obtained for different sets of nonvanishing values of the derivative couplings, meaning that their respective contributions compensate each other. Although the production may be same, the spectral distribution is certainly different and this provides a mechanism for distinguishing  between the case with vanishing derivative couplings and any other with the same total production.

The right panel of figure \ref{fig:smallmass} shows the density for small masses of the field as a function of~$\bar\gamma$ for several different constant values of $\bar\sigma$. As the values of the coupling we are considering lie well inside the   regime \eqref{eq:pertublim2}, it is not surprising that we obtain a linear behaviour of the density with the strength of the coupling $\bar\gamma$. The intensity of the linear effect of the $\bar\gamma$ term increases with increasing absolute value of the coupling constant $\bar\sigma$. It is worth noting that for positive values of $\bar\gamma$ the production diminishes, which is the opposite effect of just changing the effective mass in the de Sitter phase, giving a strong indication that the curvature effects during the transition are of vital importance for understanding the derivative couplings.
 
\begin{figure}
    \centering
    \begin{overpic}[width=.45\textwidth]{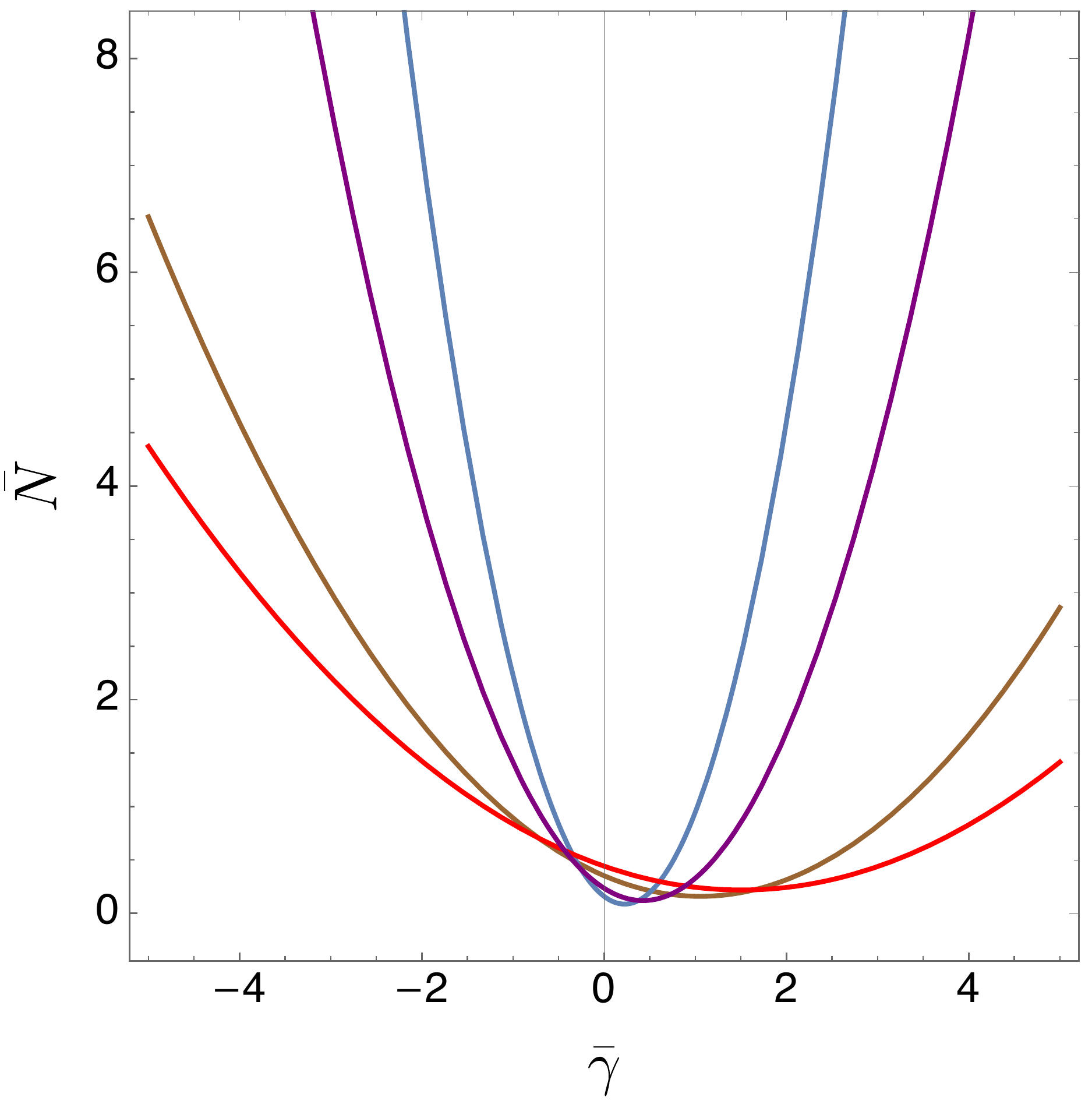}
    \put(11.5,101){{\scriptsize \sansmath $(\times 10^{-3})$}}\put(92,9){{\scriptsize \sansmath $(\times 10^{-3})$}}
    \end{overpic}
   \caption{Density of produced particles for the large mass scenario computed at $\eta_\star=50/H_0$. We show the density as a function of $\bar\gamma$ for the single value $\bar\sigma=0$ due to its negligible effect on the production (see figure \ref{fig:spectralargemass}-right). The colour code is the following: Red, $\bar m =8$; Brown, $\bar m = 10$; Purple, $\bar m=15$; Blue, $\bar m =20$.}
    \label{fig:largemass}
\end{figure}

For the large mass regime, the contribution of the $\bar\sigma$ coupling is negligible, as can be deduced from the spectral distribution (figure \ref{fig:spectralargemass}). Hence in figure \ref{fig:largemass}, we have focused only on the nonlinear dependence of the total production (for $\bar m=10$) on the $\bar\gamma$ coupling.
The production is very sensitive to the value of $\bar\gamma$. It attains a minimum value for a certain strength of the coupling. Again we see that the total production may have the same value for different $\bar\gamma$, although the spectral production is  very different. The total particle production is a quadratic function of the derivative coupling $\bar\gamma$ to very good approximation with mass dependent  coefficients.

\begin{figure}
    \centering
     \begin{overpic}[width=.45\textwidth]{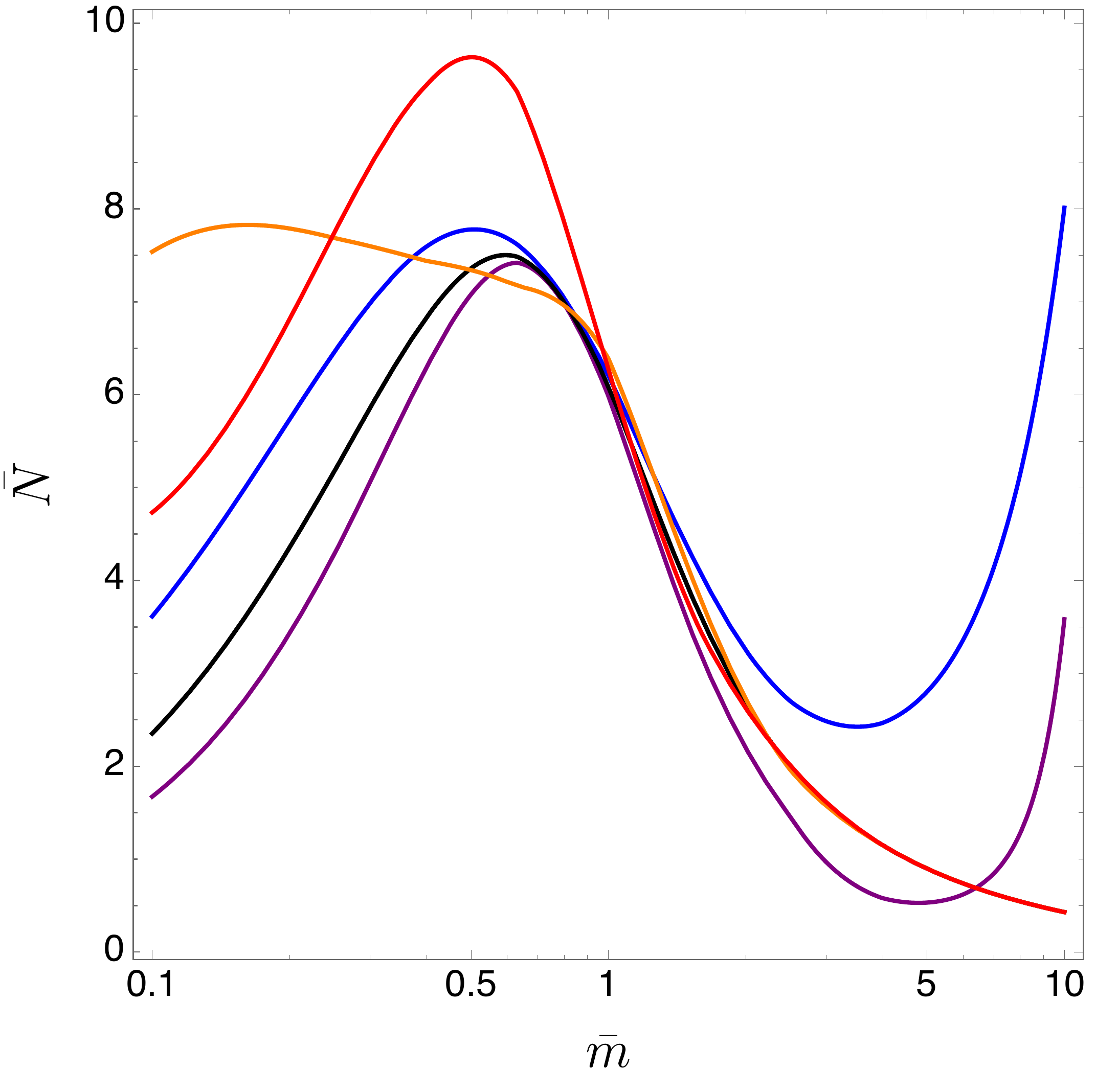}
    \put(12,101){{\scriptsize \sansmath $(\times 10^{-3})$}}\end{overpic}
   \caption{Density of produced particles computed at $\eta_\star = 50/H_0$ as a function of the mass of the field for different combinations $(\bar\gamma,\bar\sigma)$ with $\xi=1/6$ with the following colour code: Purple, $(0.01/2,0)$; Blue, $(-0.01/2,0)$; Black, $(0,0)$; Orange, $(0,1/2)$; Red, $(0,-1/2)$.}
    \label{fig:densitymass}
\end{figure}

Finally, let us see how the particle production depends on the mass of the field $\bar m$ for given values of the derivative couplings, as shown in figure \ref{fig:densitymass}. We have chosen these values so they are representative of the regimes we have studied above. More explicitly, we consider the limiting cases used in the discussions and figures of the spectral production both in $\bar\gamma$ and in~$\bar\sigma$. Smaller values of $\bar\sigma$ do not provide additional information other than inducing small perturbative changes in the total production.
We can see from figure \ref{fig:densitymass} that the two regimes we have discussed in the previous subsections are indeed representative of the behaviour for large and small masses, so we will analyse the figure in these terms.

Let us start with the small mass regime. The effect of negative $\bar\sigma$ couplings close to 1 in absolute value on the total production is less  prominent than the effect of positive values. Indeed the contribution for positive values of $\bar\gamma$ grows rapidly as the mass decreases, giving larger total production for small masses. In the rest of cases the maximum production in the small mass regime is attained for masses of order $\bar m=1$. In this region of small mass, perturbative values of $\bar\gamma$ induce linear changes in the production (as also would small values of $\bar\sigma $, although not shown in the figure) that grows in importance as the mass decreases.

The large mass regime presents some nontrivial characteristic as we have already mentioned. To begin with, the purple and blue curves describing the production for perturbative values of $\bar\gamma$ quickly grow for large masses. The contribution of positive values of $\bar\gamma$ changes from diminishing the total production for intermediate masses to wildly enhancing it after a threshold mass that depend on the specific value of the coupling. We can understand this effect using the previous analysis of the spectral production: as the mass becomes larger,  nonlinear effects enter the game. The linear effect for positive values of this coupling is to damp the amplitude of the de Sitter peak. The nonlinearity introduces a broad band of excited modes that grows with the strength of the coupling and, as we have already mentioned, becomes the principal contribution to the production. These effects are generic for any value of $\bar\gamma$.  Note that the growth in importance of the $\gamma$ coupling with larger mass is already suggested by the linear analysis of the mode as the linear contribution grows with the mass \eqref{eq:zetalargemass}. A similar behaviour could be expected for the $\bar\sigma$ coupling for the same reason. However, we  see a total production that is entirely unaffected by this coupling to the traceless Ricci curvature tensor. This negligible behaviour continues for masses greater than the ones represented. This is in agreement with the study performed in subsection \ref{sub:perturbative}, where we concluded that the $\bar\sigma$ contribution to the perturbed mode is very much suppressed (see \eqref{eq:zetalargemass}). This is also in consonance with the spectral analysis above. The production is dominated by the de Sitter phase and it is not surprising that the production decays as $\bar N\propto m^{-1}$, given the adiabatic evolution of the background geometry. 

The behaviour of the total production in the intermediate mass region is simply an interpolation of the two regimes (small and large mass) already described.

In conclusion, we have seen that including derivative couplings to the background curvature changes in a highly nontrivial way the gravitational production in the early universe for the scalar field even in the perturbative regime.

\section{Conclusions}\label{conclusions}

We have studied the role of derivative couplings to the background curvature in the gravitational production for a quantum scalar field. We have considered a coupling to the curvature scalar mediated through a coupling constant $\gamma$ and a term proportional to the traceless Ricci tensor through a constant $\sigma$. 

During the inflationary epoch of the Universe, mimicked by the de Sitter solution, only the term proportional to $\gamma$ is nonvanishing, so there is no contribution to the production due to different values of $\sigma$. We have analytically calculated both the spectral and the total production for the large mass regime, obtaining that the total production decreases with the inverse of the effective mass, which depends on the mass itself and the couplings to the curvature scalar $\xi$ and $\gamma$.

After inflation we have considered a reheating phase dominated by the oscillations of the inflaton field around the minimum of its potential. We have interpolated the behaviour of all the background quantities from their values at the end of the de Sitter phase to their values at the onset of the inflaton oscillations using sufficiently smooth functions so the results are independent of the considered order of the interpolation functions. The production in the transition phase is strongly dependent on the value of the couplings $\gamma$ and $\sigma$. To focus only on the relevance of the derivative couplings we have set $\xi=1/6$. The gravitational production quickly stabilises during the reheating phase (we are neglecting the oscillations of the background quantities during reheating).

The perturbative regime for the derivative couplings comes from requiring that the derivative terms in the action are perturbative in the sense that they are much smaller than one.   However,    small  values for the coupling constants can induce nonlinear effects because of nontrivial contributions as compared with the  field mass. In addition,  the validity of the linear approximation for the spectral production requires  additional conditions that depend on the mass.  The summary is that the perturbative regime requires that the (dimensionless) couplings  be much smaller than one, while the linear regime for the production requires conditions that depend on the mass.  In this sense, it is not surprising to have nonlinear effects in the production even though the couplings are perturbative, as happens for sufficiently large masses. 

The effects of both derivative terms are of different relative importance for a given mass of the field. Moreover, their contributions to the gravitational production depend  appreciably on the mass of the field. We have seen that for small masses,  values of $\gamma$ lying within the perturbative regime give  contributions to the particle production that are comparable to the ones due to significantly higher values of $\sigma$. In this regime, the gravitational production can change notably for perturbative values of the coupling constants so their effects cannot be neglected.

The behaviour of the production drastically changes for large masses. For values of $\gamma$ still well within the perturbative regime, the effects on the production become nonlinear while the effects of the $\sigma$ term are negligible. Hence, for large masses the production becomes very sensitive to the value of the derivative coupling $\gamma$.  This can be understood in terms of the different conditions that $\gamma$ must satisfy to remain in the linear regime and which are violated for sufficiently large masses. On the other hand note that the $\sigma$ coupling becomes irrelevant for large masses. 

To conclude, we have seen that the presence of derivative couplings to the Riemann tensor can significantly affect the gravitational production. Therefore, they can change the current picture of gravitational production of dark matter \cite{Markkanen:2015xuw, Chung:2001cb, Ema2018, Velazquez:2019mpj} as the production changes in a highly nontrivial way even when the coupling constants are in the perturbation regime. 

The derivative couplings are often neglected in all the discussions on gravitational production but, for certain theoretical field models, they can be important, and hence their gravitational production will be dominated by the effects we have discussed. This situation will be general for (pseudo-)Nambu-Goldstone bosons, whose disformal couplings dominate their phenomenology. In this context, the presence of disformal scalars, such as branons, are well-motivated from a quantum approach, and its gravitational production will differ from a traditional scalar field with nonderivative couplings.

\section*{Acknowledgments}
We wish to thank Francisco J. L\'opez Encinas for valuable discussions in the early stages of this work. This work has been financially supported in part by the MINECO (Spain) projects FIS2016-78859-P (AEI/FEDER) and FIS2017-86497-C2-2-P (with FEDER contribution). JMSV acknowledges financial support from Universidad Complutense de Madrid through the predoctoral grant CT27/16. This work was made possible by Institut Pascal at Universit\'e Paris-Saclay with the support of the P2I and SPU research departments and the P2IO Laboratory of Excellence (program Investissements davenir ANR-11-IDEX-0003-01 Paris-Saclay and ANR-10-LABX-0038), as well as the IPhT.

\bibliographystyle{JHEP}
\bibliography{references}

\providecommand{\href}[2]{#2}\begingroup\raggedright\begin{thebibliography}{10}

\bibitem{Kahlhoefer2017}
F.~Kahlhoefer, \emph{{Review of LHC Dark Matter Searches}},
  \href{https://doi.org/10.1142/S0217751X1730006X}{\emph{Int. J. Mod. Phys.}
  {\bfseries A 32} (2017) 1730006}
  [\href{https://arxiv.org/abs/1702.02430}{{\ttfamily 1702.02430}}].

\bibitem{Gaskins:2016cha}
J.~M. Gaskins, \emph{{A review of indirect searches for particle dark matter}},
  \href{https://doi.org/10.1080/00107514.2016.1175160}{\emph{Contemp. Phys.}
  {\bfseries 57} (2016) 496}
  [\href{https://arxiv.org/abs/1604.00014}{{\ttfamily 1604.00014}}].

\bibitem{Chung:1998zb}
D.~J.~H. Chung, E.~W. Kolb and A.~Riotto, \emph{{Superheavy dark matter}},
  \href{https://doi.org/10.1103/PhysRevD.59.023501}{\emph{Phys. Rev.}
  {\bfseries D 59} (1998) 023501}
  [\href{https://arxiv.org/abs/hep-ph/9802238}{{\ttfamily hep-ph/9802238}}].

\bibitem{Chung:2001cb}
D.~J.~H. Chung, P.~Crotty, E.~W. Kolb and A.~Riotto, \emph{{On the
  gravitational production of superheavy dark matter}},
  \href{https://doi.org/10.1103/PhysRevD.64.043503}{\emph{Phys. Rev.}
  {\bfseries D 64} (2001) 043503}
  [\href{https://arxiv.org/abs/hep-ph/0104100}{{\ttfamily hep-ph/0104100}}].

\bibitem{Ema2018}
Y.~Ema, K.~Nakayama and Y.~Tang, \emph{{Production of Purely Gravitational Dark
  Matter}}, \href{https://doi.org/10.1007/JHEP09(2018)135}{\emph{JHEP}
  {\bfseries 09} (2018) 135}
  [\href{https://arxiv.org/abs/1804.07471}{{\ttfamily 1804.07471}}].

\bibitem{Markkanen:2015xuw}
T.~Markkanen and S.~Nurmi, \emph{{Dark matter from gravitational particle
  production at reheating}},
  \href{https://doi.org/10.1088/1475-7516/2017/02/008}{\emph{JCAP} {\bfseries
  1702} (2017) 008} [\href{https://arxiv.org/abs/1512.07288}{{\ttfamily
  1512.07288}}].

\bibitem{Velazquez:2019mpj}
J.~A.~R. Cembranos, L.~J. Garay and J.~M. S{\'a}nchez~Vel{\'a}zquez,
  \emph{Gravitational production of scalar dark matter},
  \href{https://doi.org/10.1007/JHEP06(2020)084}{\emph{JHEP} {\bfseries 06}
  (2020) 84}.

\bibitem{Herring:2019hbe}
N.~Herring, D.~Boyanovsky and A.~R. Zentner, \emph{{Nonadiabatic cosmological
  production of ultralight dark matter}},
  \href{https://doi.org/10.1103/PhysRevD.101.083516}{\emph{Phys. Rev.}
  {\bfseries D 101} (2020) 083516}
  [\href{https://arxiv.org/abs/1912.10859}{{\ttfamily 1912.10859}}].

\bibitem{Parker:1969au}
L.~Parker, \emph{{Quantized fields and particle creation in expanding
  universes. 1.}}, \href{https://doi.org/10.1103/PhysRev.183.1057}{\emph{Phys.
  Rev.} {\bfseries 183} (1969) 1057}.

\bibitem{PhysRevD.23.347}
A.~H. Guth, \emph{Inflationary universe: A possible solution to the horizon and
  flatness problems},
  \href{https://doi.org/10.1103/PhysRevD.23.347}{\emph{Phys. Rev.} {\bfseries D
  23} (1981) 347}.

\bibitem{LINDE1983177}
A.~Linde, \emph{Chaotic inflation},
  \href{https://doi.org/https://doi.org/10.1016/0370-2693(83)90837-7}{\emph{Phys.
  Lett.} {\bfseries B 129} (1983) 177 }.

\bibitem{Liddle}
A.~Liddle and D.~Lyth, \emph{Cosmological Inflation and Large-Scale Structure}.
  Cambridge University Press, 2000.

\bibitem{Mukhanov}
V.~Mukhanov and S.~Winitzki, \emph{Introduction to Quantum Effects in Gravity}.
  Cambridge University Press, 2007.

\bibitem{Nambu:1960tm}
Y.~Nambu, \emph{Quasi-particles and gauge invariance in the theory of
  superconductivity},
  \href{https://doi.org/10.1103/PhysRev.117.648}{\emph{Phys. Rev.} {\bfseries
  117} (1960) 648}.

\bibitem{Goldstone:1961eq}
J.~Goldstone, \emph{{Field Theories with Superconductor Solutions}},
  \href{https://doi.org/10.1007/BF02812722}{\emph{Nuovo Cim.} {\bfseries 19}
  (1961) 154}.

\bibitem{Bekenstein:1992pj}
J.~D. Bekenstein, \emph{{The Relation between physical and gravitational
  geometry}}, \href{https://doi.org/10.1103/PhysRevD.48.3641}{\emph{Phys. Rev.}
  {\bfseries D 48} (1993) 3641}
  [\href{https://arxiv.org/abs/gr-qc/9211017}{{\ttfamily gr-qc/9211017}}].

\bibitem{Dobado:2000gr}
A.~Dobado and A.~L. Maroto, \emph{{The Dynamics of the Goldstone bosons on the
  brane}}, \href{https://doi.org/10.1016/S0550-3213(00)00574-5}{\emph{Nucl.
  Phys.} {\bfseries B 592} (2001) 203}
  [\href{https://arxiv.org/abs/hep-ph/0007100}{{\ttfamily hep-ph/0007100}}].

\bibitem{Cembranos:2001rp}
J.~A.~R. Cembranos, A.~Dobado and A.~L. Maroto, \emph{{Brane skyrmions and
  wrapped states}},
  \href{https://doi.org/10.1103/PhysRevD.65.026005}{\emph{Phys. Rev.}
  {\bfseries D 65} (2002) 026005}
  [\href{https://arxiv.org/abs/hep-ph/0106322}{{\ttfamily hep-ph/0106322}}].

\bibitem{Alcaraz:2002iu}
J.~Alcaraz, J.~A.~R. Cembranos, A.~Dobado and A.~L. Maroto, \emph{{Limits on
  the brane fluctuations mass and on the brane tension scale from electron
  positron colliders}},
  \href{https://doi.org/10.1103/PhysRevD.67.075010}{\emph{Phys. Rev.}
  {\bfseries D 67} (2003) 075010}
  [\href{https://arxiv.org/abs/hep-ph/0212269}{{\ttfamily hep-ph/0212269}}].

\bibitem{Cembranos:2004eb}
J.~A.~R. Cembranos, A.~Dobado and A.~L. Maroto, \emph{{Dark geometry}},
  \href{https://doi.org/10.1142/S0218271804006322}{\emph{Int. J. Mod. Phys.}
  {\bfseries D 13} (2004) 2275}
  [\href{https://arxiv.org/abs/hep-ph/0405165}{{\ttfamily hep-ph/0405165}}].

\bibitem{Cembranos:2016jun}
J.~A.~R. Cembranos and A.~L. Maroto, \emph{{Disformal scalars as dark matter
  candidates: Branon phenomenology}},
  \href{https://doi.org/10.1142/S0217751X16300155}{\emph{Int. J. Mod. Phys.}
  {\bfseries 31} (2016) 1630015}
  [\href{https://arxiv.org/abs/1602.07270}{{\ttfamily 1602.07270}}].

\bibitem{Brax:2014vva}
P.~Brax and C.~Burrage, \emph{{Constraining Disformally Coupled Scalar
  Fields}}, \href{https://doi.org/10.1103/PhysRevD.90.104009}{\emph{Phys. Rev.}
  {\bfseries D 90} (2014) 104009}
  [\href{https://arxiv.org/abs/1407.1861}{{\ttfamily 1407.1861}}].

\bibitem{Achard:2004uu}
{\scshape L3} collaboration, \emph{{Search for branons at LEP}},
  \href{https://doi.org/10.1016/j.physletb.2004.07.014}{\emph{Phys. Lett.}
  {\bfseries B 597} (2004) 145}
  [\href{https://arxiv.org/abs/hep-ex/0407017}{{\ttfamily hep-ex/0407017}}].

\bibitem{Cembranos:2011cm}
J.~A.~R. Cembranos, J.~L. Diaz-Cruz and L.~Prado, \emph{{Impact of DM direct
  searches and the LHC analyses on branon phenomenology}},
  \href{https://doi.org/10.1103/PhysRevD.84.083522}{\emph{Phys. Rev.}
  {\bfseries D 84} (2011) 083522}
  [\href{https://arxiv.org/abs/1110.0542}{{\ttfamily 1110.0542}}].

\bibitem{Cembranos:2005sr}
J.~A.~R. Cembranos, A.~Dobado and A.~L. Maroto, \emph{{Dark matter clues in the
  muon anomalous magnetic moment}},
  \href{https://doi.org/10.1103/PhysRevD.73.057303}{\emph{Phys. Rev.}
  {\bfseries D 73} (2006) 057303}
  [\href{https://arxiv.org/abs/hep-ph/0507066}{{\ttfamily hep-ph/0507066}}].

\bibitem{Cembranos:2005jc}
J.~A.~R. Cembranos, A.~Dobado and A.~L. Maroto, \emph{{Branon radiative
  corrections to collider physics and precision observables}},
  \href{https://doi.org/10.1103/PhysRevD.73.035008}{\emph{Phys. Rev.}
  {\bfseries D 73} (2006) 035008}
  [\href{https://arxiv.org/abs/hep-ph/0510399}{{\ttfamily hep-ph/0510399}}].

\bibitem{Cembranos:2003mr}
J.~A.~R. Cembranos, A.~Dobado and A.~L. Maroto, \emph{{Brane world dark
  matter}}, \href{https://doi.org/10.1103/PhysRevLett.90.241301}{\emph{Phys.
  Rev. Lett.} {\bfseries 90} (2003) 241301}
  [\href{https://arxiv.org/abs/hep-ph/0302041}{{\ttfamily hep-ph/0302041}}].

\bibitem{Cembranos:2003fu}
J.~A.~R. Cembranos, A.~Dobado and A.~L. Maroto, \emph{{Cosmological and
  astrophysical limits on brane fluctuations}},
  \href{https://doi.org/10.1103/PhysRevD.68.103505}{\emph{Phys. Rev.}
  {\bfseries D 68} (2003) 103505}
  [\href{https://arxiv.org/abs/hep-ph/0307062}{{\ttfamily hep-ph/0307062}}].

\bibitem{Maroto:2003gm}
A.~L. Maroto, \emph{{The Nature of branon dark matter}},
  \href{https://doi.org/10.1103/PhysRevD.69.043509}{\emph{Phys. Rev.}
  {\bfseries D 69} (2004) 043509}
  [\href{https://arxiv.org/abs/hep-ph/0310272}{{\ttfamily hep-ph/0310272}}].

\bibitem{Ade:2015xua}
{\scshape Planck} collaboration, \emph{{Planck 2015 results. XIII. Cosmological
  parameters}},
  \href{https://doi.org/10.1051/0004-6361/201525830}{\emph{Astron. Astrophys.}
  {\bfseries 594} (2016) A13}
  [\href{https://arxiv.org/abs/1502.01589}{{\ttfamily 1502.01589}}].

\bibitem{Mukhanov:2005sc}
V.~Mukhanov, \emph{{Physical Foundations of Cosmology}}. Cambridge University
  Press, Oxford, 2005.

\bibitem{Shale}
D.~Shale, \emph{Linear symmetries of free boson fields},
  \href{https://doi.org/10.2307/1993745}{\emph{Trans. Am. Math. Soc.}
  {\bfseries 103} (1962) 149}.

\bibitem{HoneggerRieckers}
R.~Honegger and A.~Rieckers, \emph{{Squeezing Bogoliubov transformations on the
  infinite mode CCR‐algebra}},
  \href{https://doi.org/10.1063/1.531656}{\emph{J. Math. Phys} {\bfseries 37}
  (1996) 4292}
  [\href{https://arxiv.org/abs/https://doi.org/10.1063/1.531656}{{\ttfamily
  https://doi.org/10.1063/1.531656}}].

\bibitem{Ruijsenaars}
S.~Ruijsenaars, \emph{{On Bogolyubov Transformations. 2. The General Case}},
  \href{https://doi.org/10.1016/0003-4916(78)90006-4}{\emph{Annals Phys.}
  {\bfseries 116} (1978) 105}.

\bibitem{BirrelDavies}
N.~Birrell and P.~Davies, \emph{{Quantum Fields in Curved Space}}, Cambridge
  Monographs on Mathematical Physics. Cambridge Univ. Press, Cambridge, UK, 2,
  1984,
  \href{https://doi.org/10.1017/CBO9780511622632}{10.1017/CBO9780511622632}.

\bibitem{AS}
M.~Abramowitz and I.~Stegun, \emph{Handbook of mathematical functions with
  Formulas, Graphs and Mathematical tables}. United States Department of
  Commerce. National Bureau of Standards, 1964.

\bibitem{ParkerToms}
L.~Parker and D.~Toms, \emph{Quantum Field Theory in Curved Spacetime:
  Quantized Fields and Gravity}, Cambridge Monographs on Mathematical Physics.
  Cambridge University Press, 2009.

\bibitem{PhysRevD.56.3258}
L.~Kofman, A.~Linde and A.~A. Starobinsky, \emph{Towards the theory of
  reheating after inflation},
  \href{https://doi.org/10.1103/PhysRevD.56.3258}{\emph{Phys. Rev.} {\bfseries
  D 56} (1997) 3258} [\href{https://arxiv.org/abs/hep-ph/9704452}{{\ttfamily
  hep-ph/9704452}}].

\bibitem{PhysRevD.48.647}
M.~R. de~Garcia~Maia, \emph{{Spectrum and energy density of relic gravitons in
  flat Robertson-Walker universes}},
  \href{https://doi.org/10.1103/PhysRevD.48.647}{\emph{Phys. Rev.} {\bfseries D
  48} (1993) 647}.

\end{thebibliography}\endgroup

\end{document}